\documentclass[aps,prb,reprint,citeautoscript,showpacs,floatfix,superscriptaddress]{revtex4-1}
\usepackage{bm}
\usepackage{amssymb}
\usepackage{graphicx}
\usepackage{amsmath}
\usepackage{textcomp}
\usepackage{multirow}
\usepackage{ulem}

\begin{document}

\title{Third harmonic generation of undoped graphene in Hartree-Fock approximation}

\author{J. L. Cheng}
\email{jlcheng@ciomp.ac.cn}

\affiliation{The Guo China-US Photonics Laboratory, Changchun Institute of Optics,
fine Mechanics and Physics, Chinese Academy of Sciences, 3888 Eastern
South Lake Road, Changchun, Jilin 130033, China.}
\affiliation{University of Chinese Academy of Sciences, Beijing 100049, China}
\author{J. E. Sipe}
\email{sipe@physics.utoronto.ca}

\affiliation{Department of Physics, University of Toronto, 60 St. George Street,
Toronto, Ontario M5S 1A7, Canada}
\author{Chunlei Guo}
\email{chunlei.guo@rochester.edu}

\affiliation{The Guo China-US Photonics Laboratory, Changchun Institute of Optics,
fine Mechanics and Physics, Chinese Academy of Sciences, 3888 Eastern
South Lake Road, Changchun, Jilin 130033, China.}
\affiliation{The Institute of Optics, University of Rochester, Rochester, NY 14627,
USA.}

\begin{abstract}
We theoretically investigate the effects of Coulomb interaction, at
the level of unscreened Hartree-Fock approximation, on third harmonic
generation of undoped graphene in an equation of motion framework. The unperturbed electronic states are described by a widely
used two-band tight binding model, and the Coulomb interaction is
described by the Ohno potential. The ground state is renormalized
by taking into account the
Hartree-Fock term, and the optical conductivities are obtained by
numerically solving the equations of motion. The absolute values of conductivity for
third harmonic generation depend on the photon frequency $\Omega$
as $\Omega^{-n}$ for $\hbar\Omega<1$, and then show a peak as
$3\hbar\Omega$ approaches the renormalized energy of the $M$
point. Taking 
into account the Coulomb interaction, $n$ is found to be $5.5$, which
is significantly greater than the value of $4$ found with the neglect
of the Coulomb interaction.
Therefore the Coulomb interaction enhances third harmonic generation
at low photon energies -- for  our parameters $\hbar\Omega<0.8$~eV --
and then reduces it until the photon energy reaches about
$2.1$~eV. The effect of the background dielectric constant is also
considered.
\end{abstract}
\maketitle
\section{Introduction}

The Coulomb interaction between carriers plays an important role
in determining the  band structure of a crystal and its  optical
response  \cite{Prog.QuantumElectron._30_155_2006_Kira}. In a gapped
semiconductor, the repulsive Coulomb interaction leads to the
so-called GW correction, which increases the band gap above that
obtained from the independent particle approximation
 \cite{DFT_GW_Louie1}. In contrast, the attractive Coulomb interaction,
usually between electrons and holes, leads to the formation of
excitons. Both effects must be included in a calculation to identify
the correct linear optical response  near the absorption edge of a 
semiconductor. Besides these contributions, the Coulomb interaction
also leads to scattering, and the resulting relaxation and
thermalization of carriers, beginning on a time scale of tens of
femtoseconds. For usual semiconductors, these Coulomb effects occur in
the weak interaction regime. 

For two dimensional graphene, the atomic scale thickness leads to
strong quantum confinement and reduces the Coulomb screening  \cite{Phys.Rev.B_77_081412_2008_Hwang}.
Due to the gapless linear band structure characteristic of massless
Dirac fermions, the strength of the Coulomb interaction in graphene,
described by the ratio of the Coulomb energy to the kinetic energy
 \cite{Euro.Phys.J.B_85_395_2012_Groenqvist}, is $\alpha_{g}=e^{2}/(4\pi\epsilon_{0}\epsilon\hbar v_{F})$,
assuming an effective background dielectric constant $\epsilon$ and
a Fermi velocity $v_{F}$. Taking the experimental value $v_{F}=10^{6}$~m/s,
the resulting ratio $\alpha_{g}\approx2.2/\epsilon$ indicates the
Coulomb interaction can be tuned from the weak interaction regime
($\epsilon\approx37.5$ in certain liquid environments \cite{Nanoscale_7_18015_2015_Yadav})
to the strong interaction regime ($\epsilon=1$ for a free-standing
sample). The Coulomb interaction in graphene affects its optical response
in unexpected ways. First, \textit{ab initio} calculations  \cite{Phys.Rev.Lett._103_186802_2009_Yang}
show that the GW correction cannot open the gap, but does increase
the Fermi velocity, and corrects the linear dispersion around the
Dirac points with a logarithmic function  \cite{Phys.Rev.B_84_85446_2011_Jung}.
Second, bound excitons do not exist, and excitonic effects are not
important around the Dirac points; thus, the linear optical absorption
at low photon energies is not affected by the Coulomb interaction
 \cite{Europhys.Lett._84_37001_2008_Katsnelson,Phys.Rev.Lett._103_186802_2009_Yang}.
Yet saddle point excitons can be formed around the M point of the
band structure, and the corresponding resonant optical absorption
is red-shifted by about $0.6$~eV  \cite{Phys.Rev.Lett._103_186802_2009_Yang,Nanoscale_7_18015_2015_Yadav},
with a Fano-type lineshape due to resonance with the continuum electron-hole
states  \cite{Phys.Rev.Lett._106_046401_2011_Mak,Phys.Rev.Lett._112_207401_2014_Mak}.
More generally, the optical absorption from the infrared to the visible
is found to be reduced due to the 2D Coulomb interaction, which also
provides a very fast relaxation of hot electrons  \cite{Phys.Rev.B_84_205406_2011_Malic}.

While there have been a number of investigations into the effects
of the Coulomb interaction on the linear optical response of graphene
 \cite{Phys.Rev.Lett._103_186802_2009_Yang,Phys.Rev.Lett._106_046401_2011_Mak,Phys.Rev.Lett._112_207401_2014_Mak,Nanoscale_7_18015_2015_Yadav},
there have been few theoretical studies that take into account the
effects of the Coulomb interaction on the nonlinear optical response
 \cite{Phys.Rev.B_97_115454_2018_Avetissian}. The experimental investigations
on second  \cite{Phys.Rev.Lett._122_047401_2019_Zhang} and third order
optical nonlinearities  \cite{Nat.Photon.___2018_Jiang,ACSPhoton._4_3039_2017_Alexander,Nat.Nano._X_X_2018_Soavi},
as well as studies of high harmonic generation  \cite{Science_356_736_2017_Yoshikawa,Phys.Rev.B_90_245423_2014_Al-Naib,Nat.Commun._9_1018_2018_Baudisch},
show many advantages of utilizing 2D materials in nonlinear optics
 \cite{Nat.Photon._4_611_2010_Bonaccorso,Nat.Photon._10_227_2016_Sun,Adv.Mater._30_1705963_2018_Autere}.
These include the extremely large nonlinear coefficients \cite{Europhys.Lett._79_27002_2007_Mikhailov,Phys.Rev.Lett._105_097401_2010_Hendry},
the ease of integration in photonic devices  \cite{Nat.Photon._6_554_2012_Gu,Phys.Rev.Appl._6_044006_2016_Vermeulen,ACSPhoton._4_3039_2017_Alexander,Nat.Commun._9_2675_2018_Vermeulen},
and the chemical potential tunability  \cite{Nat.Photon.___2018_Jiang,ACSPhoton._4_3039_2017_Alexander,Nat.Nano._X_X_2018_Soavi}.
Some of these features can be well predicted and understood from
calculations  within the independent particle approximation  \cite{Europhys.Lett._79_27002_2007_Mikhailov,NewJ.Phys._16_53014_2014_Cheng,*Corrigendum_NewJ.Phys._18_29501_2016_Cheng,Phys.Rev.B_91_235320_2015_Cheng,*Phys.Rev.B_93_39904_2016_Cheng,Phys.Rev.B_92_235307_2015_Cheng,Phys.Rev.B_93_085403_2016_Mikhailov}.
Because the Coulomb interaction hardly affects the linear optical
absorption, it might be  natural to assume that its effects on the
nonlinear optical response would also be small. However, a recent study
by Avetissian and Mkrtchian \cite{Phys.Rev.B_97_115454_2018_Avetissian}
reported a large enhancement of harmonic generation at THz frequencies
due to the 2D Coulomb interaction. In the present work, we model the Coulomb
interaction by the Ohno potential and consider its effects on third
harmonic generation, in the unscreened Hartree-Fock approximation,
for fundamental photon energies between $0.2$ eV and $3.5$ eV. The
effects of a  background dielectric constant are also considered.

We organize this paper as following. In Sec.~\ref{sec:model} we
set up the equation of motion in the Hartree-Fock approximation. In
Sec.~\ref{sec:res} we present our numerical scheme and numerical
results for the band structure, the density of states, the linear
conductivity, and the nonlinear conductivity for third harmonic generation;
the effect of the background dielectric constant is also investigated.
In Sec.~\ref{sec:con} we conclude.

\section{Model\label{sec:model}}

We describe the dynamics of the electrons by a density matrix with
components $\rho_{\alpha\beta\bm{k}}(t)$, where $\alpha$ and $\beta$ label
the atom sites $A$ or $B$, and $\bm{k}$ is a crystal wave vector.
With the application of an electric field $\bm{E}(t)$, the density
matrix components satisfy the equation of motion 
\begin{align}
i\hbar\partial_{t}{\rho}_{\alpha\beta\bm{k}}(t) & =[{H}_{\bm{k}+e\bm{A}(t)/\hbar}^{0}+{H}_{\bm{k}}^{\text{HF}}(t),{\rho}_{\bm{k}}(t)]_{\alpha\beta}\nonumber \\
 & +e\bm{E}(t)\cdot(\bm{\tau}_{\alpha}-\bm{\tau}_{\beta}){\rho}_{\alpha\beta\bm{k}}(t)\nonumber \\
 & -i\Gamma[{\rho}_{\alpha\beta\bm{k}}(t)-{\rho}_{\alpha\beta(\bm{k}+e\bm{A}(t)/\hbar)}^{0}]\,.\label{eq:neweom}
\end{align}
Here $H_{\bm{k}}^{0}$ is a tight binding Hamiltonian  \cite{NewJ.Phys._16_53014_2014_Cheng},
formed by the $p_{z}$ orbitals of carbon atoms, 
\begin{align}
{H}_{\alpha\beta\bm{k}}^{0}=-\gamma_{0}\left(f_{\bm{k}}\delta_{\alpha,A}\delta_{\beta,B}+f_{\bm{k}}^{\ast}\delta_{\alpha,B}\delta_{\beta,A}\right)\,,
\end{align}
where $\gamma_{0}$ is the nearest neighbour hopping parameter and
$f_{\bm{k}}=1+e^{-i\bm{k}\cdot\bm{a}_{1}}+e^{-i\bm{k}\cdot\bm{a}_{2}}$
is the structure factor, with $\bm{a}_{i}$ the primitive lattice
vectors and $\bm{\tau}_{\alpha}$ the displacement of the $\alpha$th
atom in the unit cell. The last term in Eq.~(\ref{eq:neweom})
describes the relaxation processes phenomenologically, with a relaxation
parameter $\Gamma$; the system relaxes to an equilibrium state $\rho_{\bm{k}}^{0}$
in the moving frame, as will be discussed in detail below. In Appendix~\ref{app:derivation},
we give a brief derivation of Eq.~(\ref{eq:neweom}) based on the
tight binding model.

The carrier-carrier Coulomb interaction is included in the unscreened
Hartree-Fock approximation through the term ${H}_{\alpha\beta\bm{k}}^{\text{HF}}(t)$,
which is given by 
\begin{align}
{H}_{\alpha\beta\bm{k}}^{\text{HF}}(t) & =-\int d\bm{k}^{\prime}V_{\alpha\beta(\bm{k}-\bm{k}^{\prime})}{\rho}_{\alpha\beta\bm{k}^{\prime}}(t)\,.
\end{align}
The Coulomb interaction term $V_{\alpha\beta\bm{k}}=\sum_{j}e^{i\bm{k}\cdot\bm{R}_{j}}V_{j;\alpha\beta}$
is taken to be the Fourier transform of the Ohno potential \cite{Phys.Rev.B_75_035407_2007_Jiang}
\begin{align}
V_{i,\alpha\beta} & =\frac{U}{\epsilon\sqrt{\left(\dfrac{4\pi\epsilon_{0}|\bm{R}_{i}+\bm{\tau}_{\alpha}-\bm{\tau}_{\beta}|U}{e^{2}}\right)^{2}+1}}\,,
\end{align}
where $U$ is an onsite energy, $\epsilon$ can be considered as an
effective background dielectric constant \footnote{For graphene
  embedded inside 
  two different materials, the background 
dielectric constant is the average of the dielectric constants. Because
the interaction is mainly through the electric force outside of the
graphene plane, the screening from the graphene electrons is ignored.}, and the $\bm{R}_{i}$ are lattice vectors. With this  Ohno potential
the interaction between the A and B sites is also considered, as opposed
to  the widely used Coulomb potential in a continuum model, such as
based on a $\bm{k}\cdot\bm{p}$ Hamiltonian, where the sites are treated
with a  pseudospin description  \cite{Phys.Rev.B_97_115454_2018_Avetissian}.
 We separate the HF term into two contributions, 
\begin{align}
{H}_{\alpha\beta\bm{k}}^{\text{HF}}(t) & =\lambda_{m}{H}_{\alpha\beta(\bm{k}+e\bm{A}(t)/\hbar)}^{\text{HF};(0)}+\lambda_{e}{H}_{\alpha\beta\bm{k}}^{\text{HF};(1)}(t)\,,\\
{H}_{\alpha\beta\bm{k}}^{\text{HF};(0)} & =-\int d\bm{k}^{\prime}V_{\alpha\beta(\bm{k}-\bm{k}^{\prime})}{\rho}_{\alpha\beta\bm{k}^{\prime}}^{0}\,,\label{eq:nsc1}
\end{align}
with two auxiliary parameters $\lambda_{m}=\lambda_{e}=1$. The first
term exists even in the absence of external field. It takes into account
the Coulomb interaction at the level of a mean field approximation
(MFA), and it can modify the band structure significantly \cite{Phys.Rev.Lett._99_226801_2007_Hwang,Phys.Rev.Lett._103_186802_2009_Yang}.
Accordingly, it also affects the equilibrium distribution. The second
term, ${H}_{\alpha\beta\bm{k}}^{\text{HF};(1)}(t)={H}_{\alpha\beta\bm{k}}^{\text{HF}}(t)-{H}_{\alpha\beta(\bm{k}+e\bm{A}(t)/\hbar)}^{\text{HF};(0)}$,
describes the interaction between optically excited carriers. To better
understand the two contributions, $\lambda_{m/e}$ will be
intentionally changed in our numerical calculation to include ($=1$)
or exclude ($=0$) an effect.

Before we can solve Eq.~(\ref{eq:neweom}), it is important to determine
$\rho_{\bm{k}}^{0}$, the ground state. While the possibility
of  forming a new excitonic ground state in the strong interaction
regime has been extensively discussed \cite{Euro.Phys.J.B_85_395_2012_Groenqvist,J.Opt.Soc.Am.B_29_A86_2012_Stroucken,Phys.Rev.B_84_205445_2011_Stroucken},
for most   experimental scenarios  there is a  SiO$_{2}$ or Si substrate,
with a dielectric constant larger than 2; thus the Coulomb interaction
should be in the weak interaction regime. Therefore  we limit ourselves
to the single-particle ground state. In our treatment, the Coulomb
interaction has two main consequences: First, it  modifies the band
structure through ${H}_{\alpha\beta\bm{k}}^{\text{HF};(0)}$. The
eigen energies $\varepsilon_{s\bm{k}}$ and eigenstates $C_{s\bm{k}}=\begin{pmatrix}C_{s\bm{k}}^{A}\\
C_{s\bm{k}}^{B}
\end{pmatrix}$ with a band index $s=\pm$ are determined from the Schr\"odinger equation
\begin{align}
\sum_{\beta}\left(H_{\alpha\beta\bm{k}}^{0}+{H}_{\alpha\beta\bm{k}}^{\text{HF};(0)}\right)C_{s\bm{k}}^{\beta}=\varepsilon_{s\bm{k}}C_{s\bm{k}}^{\alpha},\label{eq:nsc2}
\end{align}
and the  equilibrium density matrix is calculated from 
\begin{align}
{\rho}_{\alpha\beta\bm{k}}^{0} & =\sum_{s}\frac{C_{s\bm{k}}^{\alpha}\left(C_{s\bm{k}}^{\beta}\right)^{\ast}}{1+e^{(\varepsilon_{s\bm{k}}-\mu)/(k_{B}T)}}\label{eq:nsc3}
\end{align}
for a specified  chemical potential $\mu$ and temperature $T$. Equations~(\ref{eq:nsc1}),
(\ref{eq:nsc2}), and (\ref{eq:nsc3}) form a self-consistent set
of equations, and they are solved iterately. By initially setting
${H}_{\alpha\beta\bm{k}}^{\text{HF};(0)}=0$ in Eq.~(\ref{eq:nsc2}),
we get $C_{s\bm{k}}^{\beta}$; next we calculate $\rho_{\bm{k}}^{0}$
using Eq.~(\ref{eq:nsc3}), and then we update ${H}_{\alpha\beta\bm{k}}^{\text{HF};(0)}$
using Eq.~(\ref{eq:nsc1}), repeating this procedure until $\rho_{\alpha\beta\bm{k}}^{0}$
is converged.  A second consequence of the Coulomb interaction is
that the term ${H}_{\alpha\beta\bm{k}}^{\text{HF};(1)}(t)$ leads
to  the excitonic effects (EE) in the framework of carrier dynamics
 \cite{QuantumKineticsinTransportandOpticsofSemiconductors1}.
\begin{widetext}
\begin{figure*}[!htbp]
  \centering
    \includegraphics[height=5.3cm]{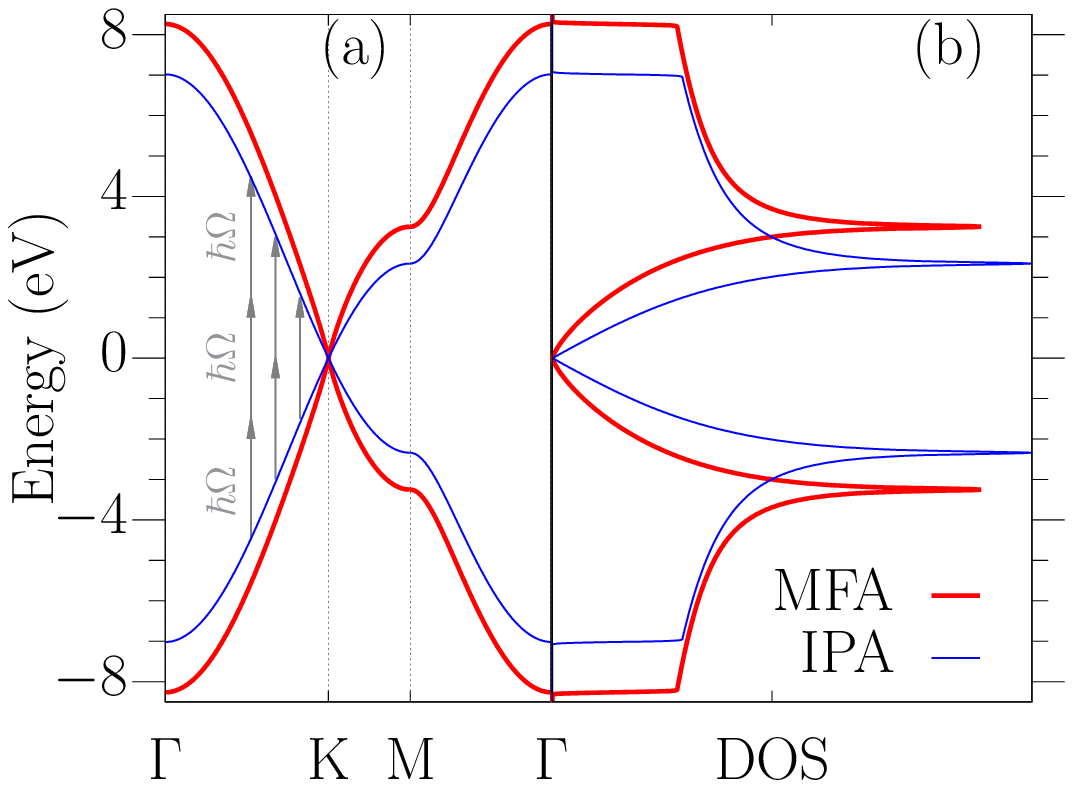}
    \includegraphics[height=5.3cm]{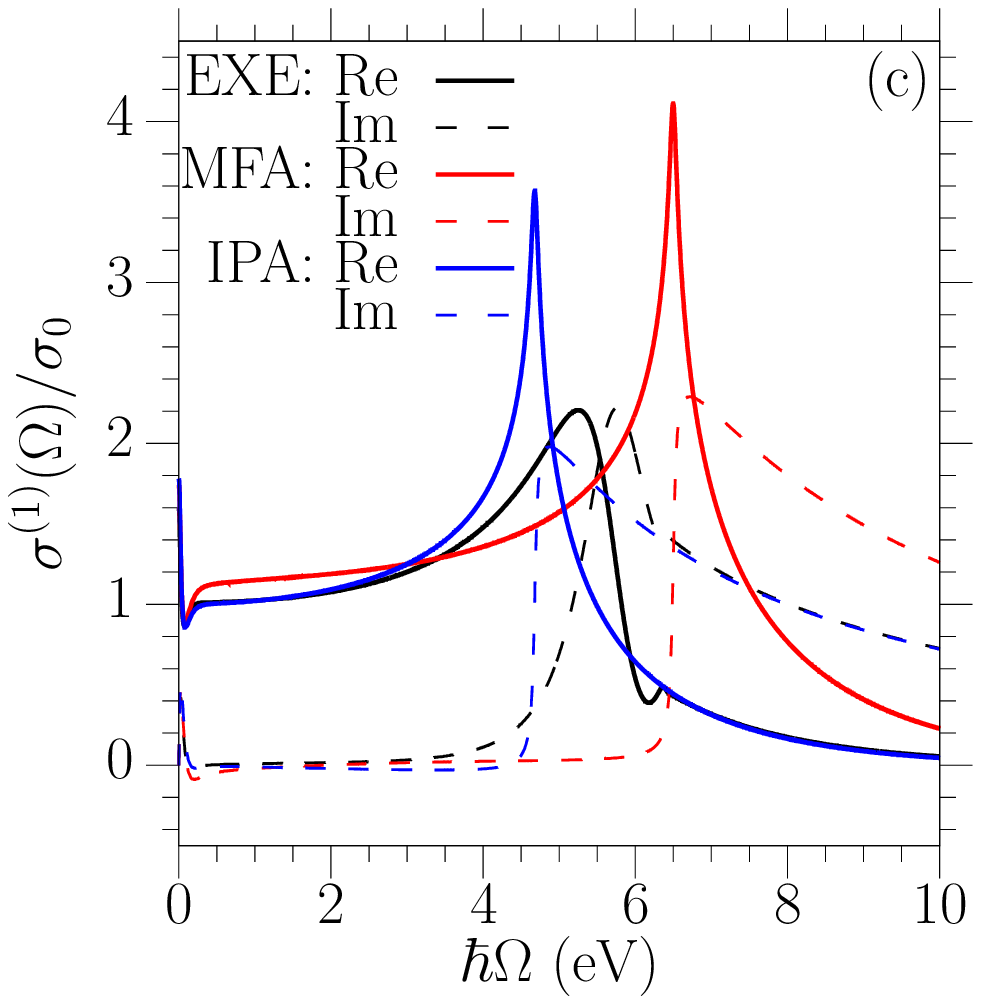}
    \includegraphics[height=5.3cm]{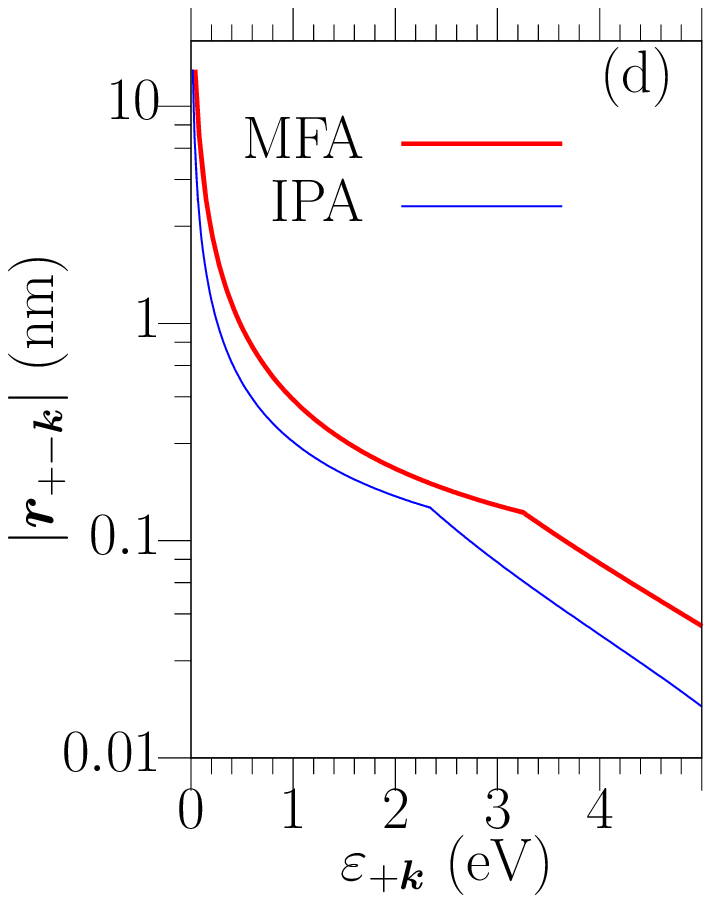}
\caption{Effects of HF term on (a) the band structure, (b) the density of states,
(c) the linear optical conductivity, and (d) the energy dependence
of the dipole transition matrix elements between the two bands along
the high symmetry lines K$\to$M$\to\Gamma$. The gray arrows in (a)
indicate the resonant transitions by multiple photons. }
\label{fig:linear}
\end{figure*}
\end{widetext}

In this paper the main quantity of interest is the  current density,
which is given by $\bm{J}(t)=(2\pi)^{-2}\int
d\boldsymbol{k}\bm{J}_{\bm{k}}(t)$ with 
\begin{align}
\bm{J}_{\bm{k}}(t) & =-e\sum_{\alpha\beta}{\bm{v}}_{\beta\alpha\bm{k}}(t){\rho}_{\alpha\beta\bm{k}}(t)\,,\\
{\bm{v}}_{\alpha\beta\bm{k}}(t) & =\frac{1}{i\hbar}\left\{ (\bm{\tau}_{\alpha}-\bm{\tau}_{\beta}+i\bm{\nabla}_{\bm{k}})\left[{H}_{\alpha\beta(\bm{k}+e\bm{A}(t)/\hbar)}^{0}\right.\right.\nonumber \\
 & \left.\left.+{H}_{\alpha\beta\bm{k}}^{\text{HF}}(t)\right]\right\} \,.
\end{align}

\section{Results\label{sec:res}}
In our numerical calculation, we adopt the parameters $\gamma_{0}=2.34$~eV,
$U=8.29$~eV, $\hbar/\Gamma=20$~fs, $\mu=0$~eV, $T=300$~K, 
and divide the Brillouin  zone into a $N\times N$ grid with $N=1500$;
our main results are not very sensitive to the  exact values of $\gamma_{0}$
and $U$. The time differential is discretized by a fourth-order Runge-Kutta
method with a time step $0.05$~fs. We consider background dielectric
constants $\epsilon$ varying between $2$ and $9$. The results are
presented for three different approaches: the independent particle
approximation (IPA, with $\lambda_{m}=\lambda_{e}=0$), the mean field
approximation (MFA, with $\lambda_{m}=1$ and $\lambda_{e}=0$), and
taking into account excitonic effects (EXE, with $\lambda_{m}=\lambda_{e}=1$).

We numerically calculate the current density for a periodic field
$\bm{E}(t)=E_{0}\hat{\bm{x}}e^{-i\Omega t}+c.c.$, with $E_{0}=5\times10^{6}$~V/m,
in the time range $0\le t\le300$~fs \footnote{This work considers the third harmonic generation in the weak
field regime, thus the use of a periodic field follows the standard
perturbative treatment, which should be suitable in experiments where
the pulse duration is much longer than the relaxation time \cite{Phys.Rev.B_92_235307_2015_Cheng}.}. Because the relaxation time is much shorter than $300$~fs, we
can approximate the current density in the last period as $J_{x}(t)=\sum_{n}\left[J_{x}^{(n)}e^{-in\Omega t}+c.c\right]$.
The linear conductivity can be obtained by $\sigma^{(1)}(\Omega)=J_{x}^{(1)}/E_{0}$,
and the THG conductivity is obtained by $\sigma^{(3)}(\Omega)=J_{x}^{(3)}/E_{0}^{3}$.
For THG, the photon energies is chosen in the range of $0.2\le\hbar\Omega\le3.4$~eV.

Insight into the dynamics induced by the applied field  can be gained
by looking at  the transition energy resolved conductivities, which
are defined as 
\begin{align}
\sigma^{(n)}(\Omega,\epsilon_{t}) & =\int\frac{d\bm{k}}{(2\pi)^{2}}\frac{J_{x;\bm{k}}^{(n)}}{E_{0}^{n}}\delta(\epsilon_{t}-(\varepsilon_{+\bm{k}}-\varepsilon_{-\bm{k}}))\,,
\end{align}
where $J_{x;\bm{k}}^{(n)}$ is the $n$th order Fourier transformation
of $J_{x;\bm{k}}(t)$. They describe how the electronic states at
a given transition energy $\epsilon_{t}$ contribute to the optical
conductivity at a fundamental frequency $\Omega$. In the IPA and
MFA approximations,  $|\sigma^{(n)}(\Omega,\epsilon_{t})|$ exhibits
a  peak at  $\epsilon_{t}=m\hbar\Omega$ for the $m$-photon resonant
optical transition for $1\le m\le n$, as would be expected within a single-particle
description. In Fig.~\ref{fig:linear} (a) we illustrate the electronic
states for $m$-photon resonant transitions at $m=1,2,3$.  With the inclusion of HF term, the energy shift of these
peaks can be used to identify excitonic effects. Very generally, the
conductivities are given by the integral over transition energies
of these resolved quantities,  $\sigma^{(n)}(\Omega)=\int d\epsilon_{t}\sigma^{(n)}(\Omega,\epsilon_{t})$.

\subsection{Band structure and linear conductivity}
Figure~\ref{fig:linear} (a) and (b) give the single particle band
structure and density of states (DOS) in the IPA and the MFA for $\epsilon=3$.
In the IPA, the dispersion is linear around the Dirac points with
a Fermi velocity $v_{F}=\sqrt{3}a_{0}\gamma_{0}/2\hbar\approx0.76\times10^{6}$~m/s,
and the DOS is also approximately linear in a large energy range between
about $-1.5$~eV and $1.5$~eV. There are significant changes in
the band structure as we move to the MFA. Our results are in line
with a host of other calculations showing that  the single particle
bands deviate from linear dispersion  \cite{Phys.Rev.Lett._99_226801_2007_Hwang}. 
And even away from the Dirac points the Fermi velocity increases.
The band energy at the M point increases from $2.34$~eV in the IPA
to $3.25$~eV in the MFA, and the energy at the $\Gamma$ point also
increases from $7$~eV in the IPA to $8.27$~eV in the MFA. The
increase of these characteristic \begin{widetext}
\begin{figure*}[!htp]
  \centering
    \includegraphics[height=5.cm]{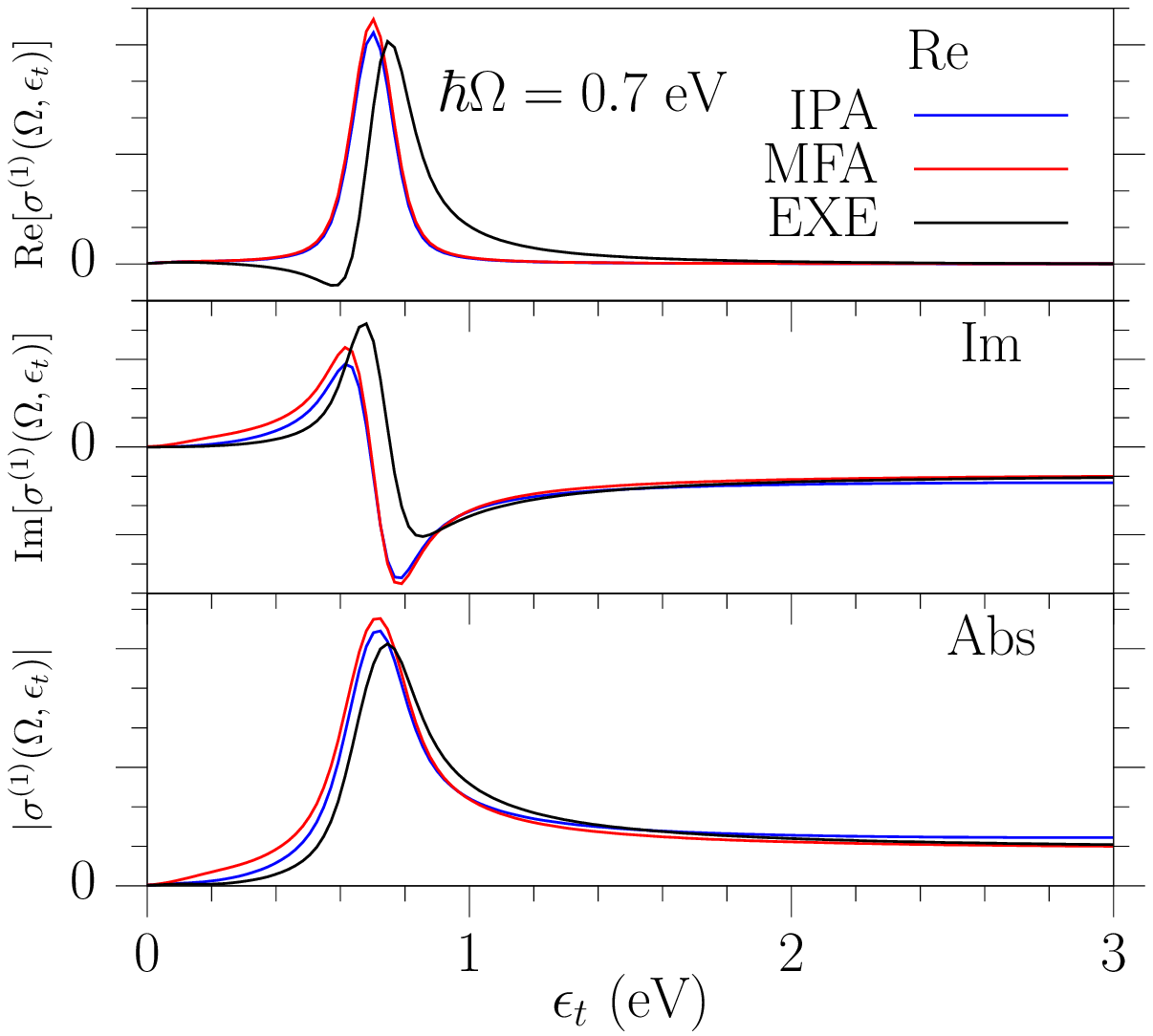}
  \includegraphics[height=5.cm]{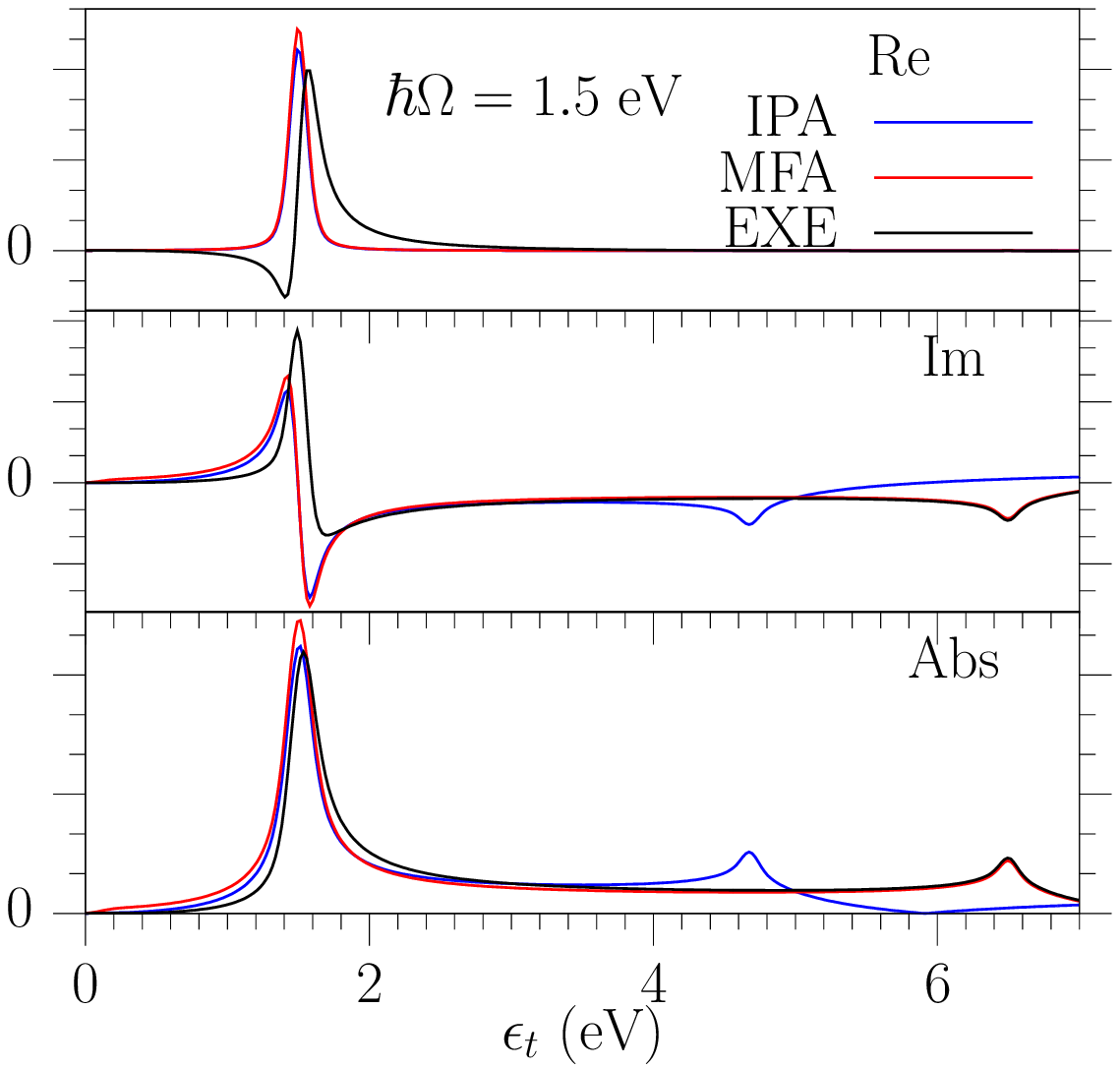}
  \includegraphics[height=5.cm]{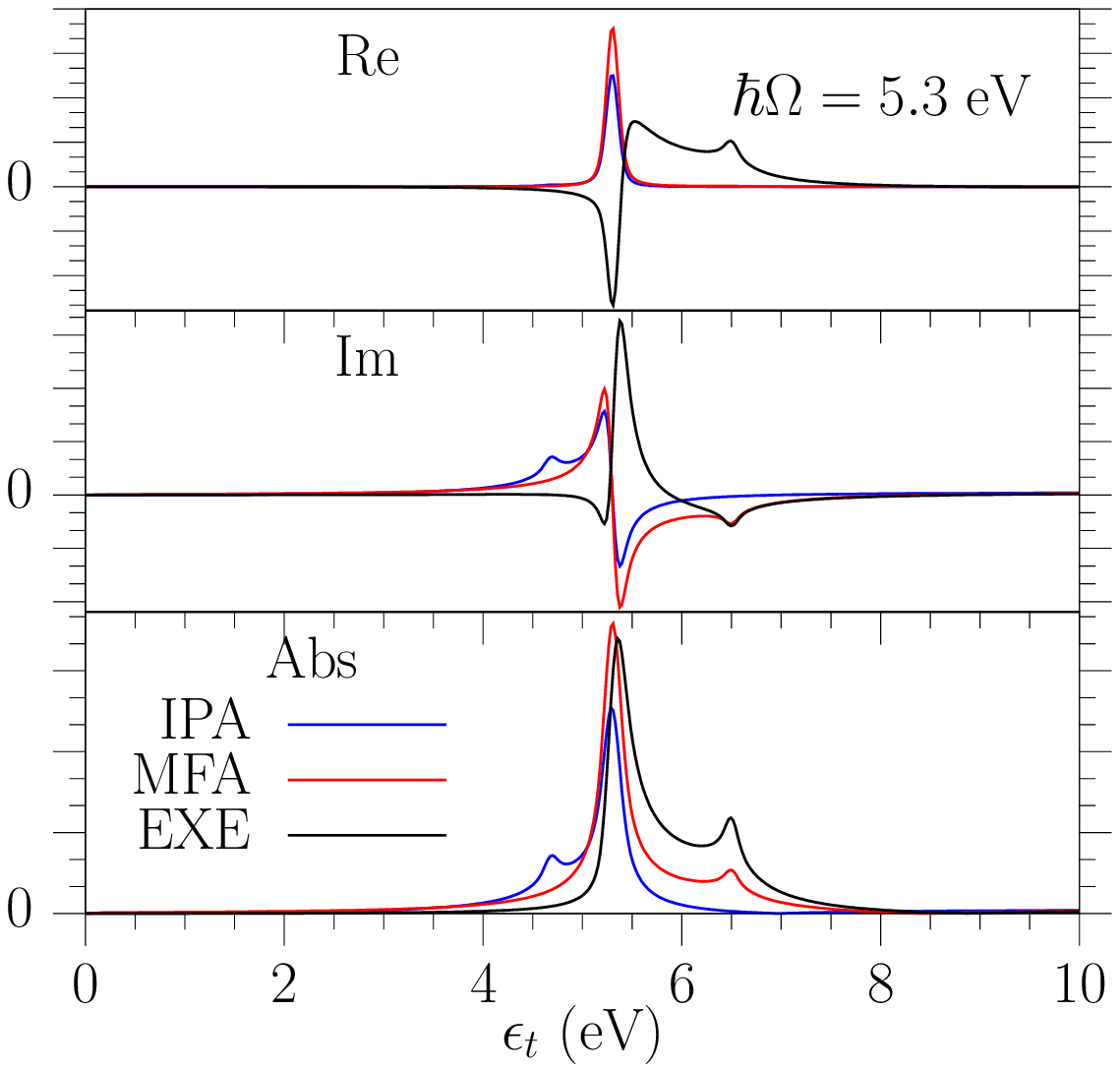}
\caption{Transition energy resolved linear conductivity for different excitation
frequency $\hbar\Omega=0.7$, $1.5$, and $5.3$~eV in the IPA, MFA, and 
EXE models. The values for the real part, the imaginary part, and
the absolute values are presented.}
\label{fig:linearspw} 
\end{figure*}
\end{widetext}energies by the HF term shows 
behavior similar to 
the effects of GW corrections  in gapped semiconductors.
Although both the IPA and MFA are single-particle  approximations,
their different band structures result in 
 a clear difference between
their predicted linear optical conductivities, 
as shown in Fig.~\ref{fig:linear}(c). The MFA corrections to the
dispersion also enhance the real parts of the linear conductivity
for photon energies less than $2.5$~eV. The enhancement is a joint
effect involving the  enhanced  dipole matrix elements shown in
Fig.~\ref{fig:linear} (d) and the decreased DOS shown in
Fig.~\ref{fig:linear} (b), and the former dominates.
The result shows how the MFA modifies the single particle electronic states. When the photon energy matches the optical transition energy
at the $M$ point, the real part of the conductivity shows a very
sharp peak in both the IPA and MFA, induced by the van Hove singularity
at the $M$ point. With the further inclusion of excitonic effects,
the resulting linear conductivity in the EXE exhibits three main features:
(1) The conductivity for $\hbar\omega<3$~eV is very close to the
universal conductivity obtained in the IPA. (2) The singularity peak
is shifted from a photon energy $6.5$~eV to $5.3$~eV, which indicates
a strong excitonic effect around the $M$ point. The exciton binding
energy is about $1.2$~eV, of the same order of magnitude for excitons
in other 2D materials. (3) The broadened peak at the $M$ point, followed
by a small dip, confirms the Fano resonance between the saddle excitons
and continuum electron-hole states. These results  agree qualitatively
with an \textit{ab initio} calculation  \cite{Phys.Rev.Lett._103_186802_2009_Yang},
which provides a good check of the reasonableness of our model. Roughly
speaking, the linearity conductivity plot for EXE lies between that
of IPA and MFA, indicating an interference between the mean field
and excitonic contributions. We will see this kind of ``undoing''
of the mean field contributions by the excitonic contributions is
even more pronounced when we consider the nonlinear response below.

In Fig.~\ref{fig:linearspw} we plot the transition energy resolved
conductivity, $\sigma^{(1)}(\Omega,\epsilon_{t})$, for the excitation
photon energies $\hbar\Omega=0.7$, $1.5$, and $5.3$~eV. For the
single particle models (the IPA and MFA), a Lorentzian shape $\sigma^{(1)}(\Omega,\epsilon_{t})\propto i\left(\hbar\Omega-\epsilon_{t}+i\Gamma\right)^{-1}$
results. The real part of $\sigma^{(1)}(\Omega,\epsilon_{t})$, which
corresponds to the absorption at each transition energy $\epsilon_{t}$,
is always positive and localized around $\epsilon_{t}\sim\hbar\Omega$
with a broadening approximately determined by the relaxation parameter;
this is as expected for a single particle theory. The difference between
the IPA and the MFA is mainly due to  the different optical transition
matrix elements. For the imaginary part, the band structure and the
density of states play important roles, with peaks around the singularity
at the $M$ points ($\epsilon_{t}\approx4.6$~eV for the IPA and
6.5~eV for the MFA). With the inclusion of the excitonic effects,
$\sigma^{(1)}(\Omega,\epsilon_{t})$ deviates  from its MFA behavior
only around the peak at $\epsilon_{t}\approx\hbar\Omega$. For $\hbar\Omega=0.7$
and $1.5$~eV, the absolute value only changes slightly, but the
mixture of the real and imaginary parts indicates a  phase change
of the optical transition matrix elements. Overall, excitonic effects
are very weak for low photon energies, agreeing with the $ab$ $initio$
results. For $\hbar\Omega=5.3$~eV, the absolute value of $\sigma^{(1)}(\Omega,\epsilon_{t})$
include a significant contribution from the electronic states at the
$M$ point for  $\epsilon_{t}>\hbar\Omega$, showing the  important
role played by  excitonic effects.

\subsection{Third harmonic generation}

\begin{figure}[!htpb]
\centering \includegraphics[height=7.cm]{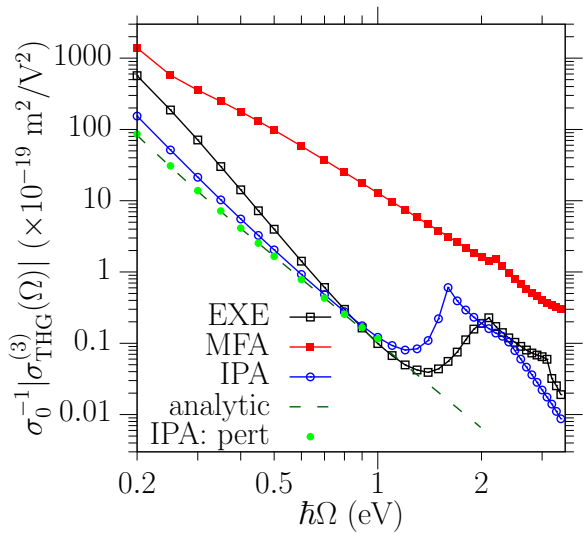}
\caption{Effects of HF term on the nonlinear conductivity for third harmonic
generation. The dotted curve gives the perturbative conductivity obtained
by an analytic expression \cite{Phys.Rev.B_91_235320_2015_Cheng,*Phys.Rev.B_93_39904_2016_Cheng};
all the other three curves are obtained under a field $E_{0}=5\times10^{6}$~V/m.
The blue (red) curves are the results obtained from the single particle
band structure with (without) HF corrections, and the black curves
are obtained with the excitonic effects. The green dots are obtained
using a field $E_{0}=10^{6}$~V/m.}
\label{fig:thg} 
\end{figure}
We now turn to THG, the conductivities of which are plotted in Fig.~\ref{fig:thg}
for IPA, MFA, and EXE. We first look at the spectra in the IPA. The
absolute value of conductivity decreases with the photon energy, approximately
following  a power law $\propto\Omega^{-n_{0}}$ with $n_{0}\sim4.4$,
and reaches a minimum at $\hbar\Omega\approx1.2$~eV; then it increases
to a maximum at $\hbar\Omega\sim1.6$~eV, and decreases again afterwards.
This spectrum is very similar to  perturbative results in literature
 \cite{NewJ.Phys._16_53014_2014_Cheng,Phys.Rev.B_96_035206_2017_Liu}.
Ignoring relaxation, the analytic perturbative conductivity  \cite{Phys.Rev.B_91_235320_2015_Cheng,*Phys.Rev.B_93_39904_2016_Cheng}
gives $\sigma_{\text{THG}}\propto\Omega^{-4}$ for low photon energies,
shown in Fig.~\ref{fig:thg} as a dashed curve. Our numerical results
give a faster decay ($n_{0}>4$) because the simulation field strength
$E_{0}=5\times10^{6}$~V/m is slightly beyond the perturbative limit.
Therefore saturation effects start to play a role, and they enhance
the THG due to the extra optically excited carriers from the
one-photon absorption  \cite{Phys.Rev.B_92_235307_2015_Cheng}. 
 As might be expected, the saturation is more important at low photon
energy, where the saturation intensity is lower. The perturbative
limit is obtained for a weaker field $E_{0}=10^{6}$~V/m, as shown
by the green filled dots in Fig.~\ref{fig:thg}, which agree with
the analytic results very well. Because a much longer simulation time
for this weak field is required, we keep our other calculations using
$E_{0}=5\times10^{6}$~V/m. The peak around $1.6$~eV is induced
by the three photon resonant transition at the van Hove singularity
point. This is consistent with an analytic perturbative result  \cite{Phys.Rev.B_96_035206_2017_Liu}.

When the Coulomb interaction is included at the level of MFA, the
spectra are obviously different, shown as filled red squares in Fig.~\ref{fig:thg}.
For the calculated photon energies, the values in the MFA are orders
of magnitude larger than those in the IPA, mirroring the change in
the linear conductivity as we move from IPA to MFA, although the increase
is much greater here.  Except for a very weak peak around $\hbar\Omega\approx2.2$~eV,
which corresponds to the three photon resonant transition at the $M$
point, the whole spectrum follows a power law $\propto\Omega^{-n_{m}}$
with $n_{m}\approx3$ for $\hbar\Omega<3.4$~eV. When the excitonic
effects are included, the THG conductivity again changes dramatically,
as shown in black square in Fig.~\ref{fig:thg}. At low photon energies,
$\hbar\Omega<1$~eV, the THG shows a different scaling with $\Omega$
than seen in either the IPA or the MFA, scaling as $\Omega^{-n_{e}}$
with $n_{e}\sim5.5$; compared to the IPA results, the  EXE are
 enhanced by about 3 times at $\hbar\Omega=0.2$~eV, but reduced
about 20\% at $\hbar\Omega=1$~eV. When the photon energy increases,
a peak appears at $\hbar\omega\approx2.1$~eV, which is the same
energy for the weak peak appearing in MFA. In brief, the Coulomb interaction
affects THG mainly in the following aspects: (1) THG in MFA is orders
of magnitude larger than that in IPA. The increase is even much larger
than that calculated for the linear response. Therefore the details
of the band structure are very important for understanding the optical
nonlinearity. (2) The further inclusion of EXE can bring the THG very
close to the results in IPA, showing a strong interference between
the mean field contribution and the excitonic contribution. (3) At low
photon energies, the Coulomb 
interaction changes the photon energy dependence, with the power index
changing from $n_{0}=4.4$ in the IPA to $n_{m}=3$ in the MFA and
$n_{e}=5.5$ in the EXE. (4) For the  EXE spectra, the three-photon
resonance with the $M$ point gives a peak at $\hbar\Omega\approx2.1$~eV.
This energy does not correspond to the one third of the $M$ point
exciton energy (~$5.3\text{~eV}/3\sim1.8$~eV), but is very close
to the one third of the $M$ point energy in the
MFA~($6.5\text{~eV}/3\sim2.16$~eV). The lack of an energy shift may indicate that the excitonic effects are not important
for the three-photon resonance with the $M$ point. These four features
indicate the importance of the Coulomb interaction on the THG in an
intrinsic graphenem, where both (1)
and (2) are very similar to the effects of Coulomb interaction on
the linear optical response.

\begin{widetext}
  \begin{figure*}[!htp]    
    \centering
  \includegraphics[height=5cm]{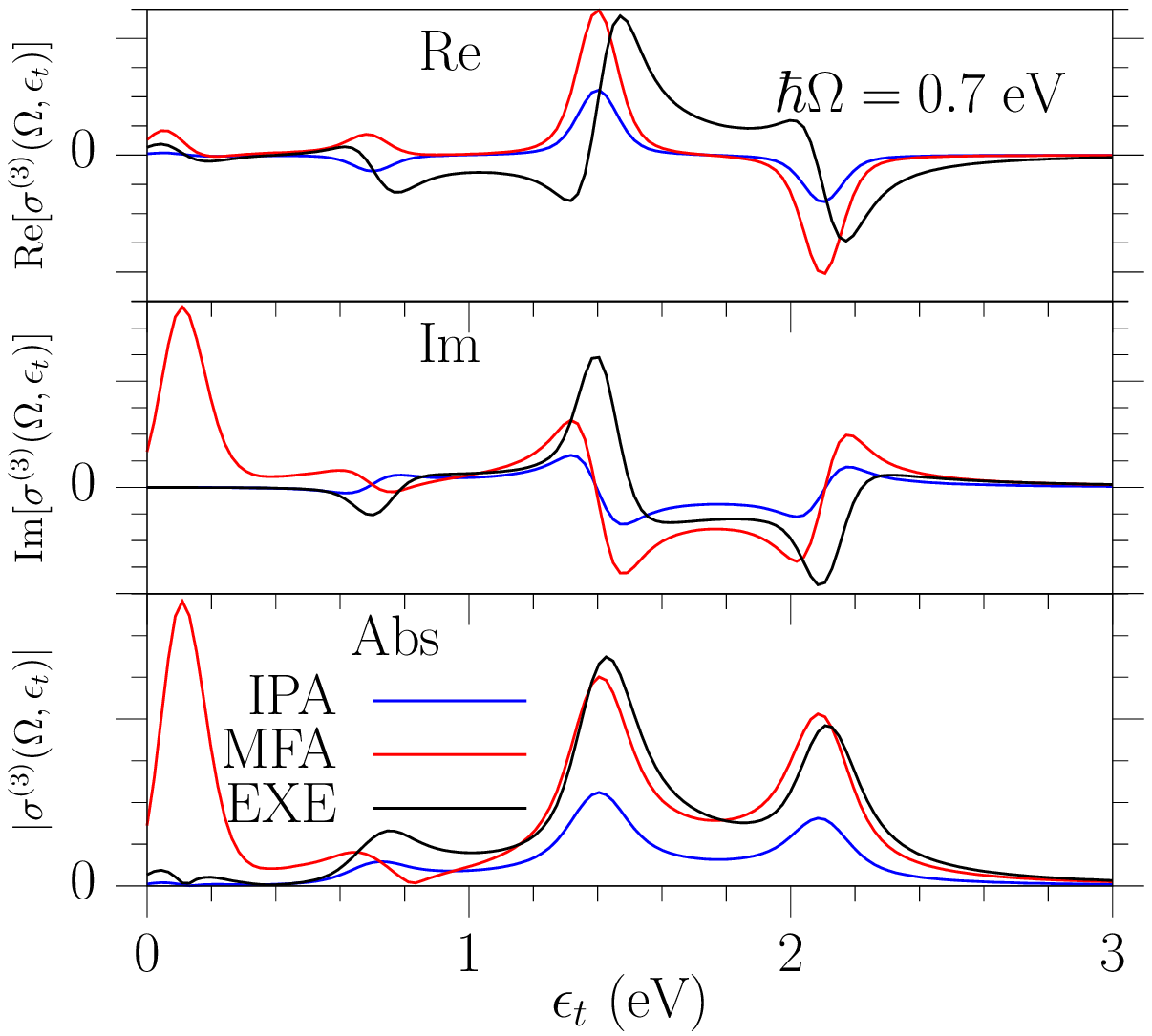}
  \includegraphics[height=5cm]{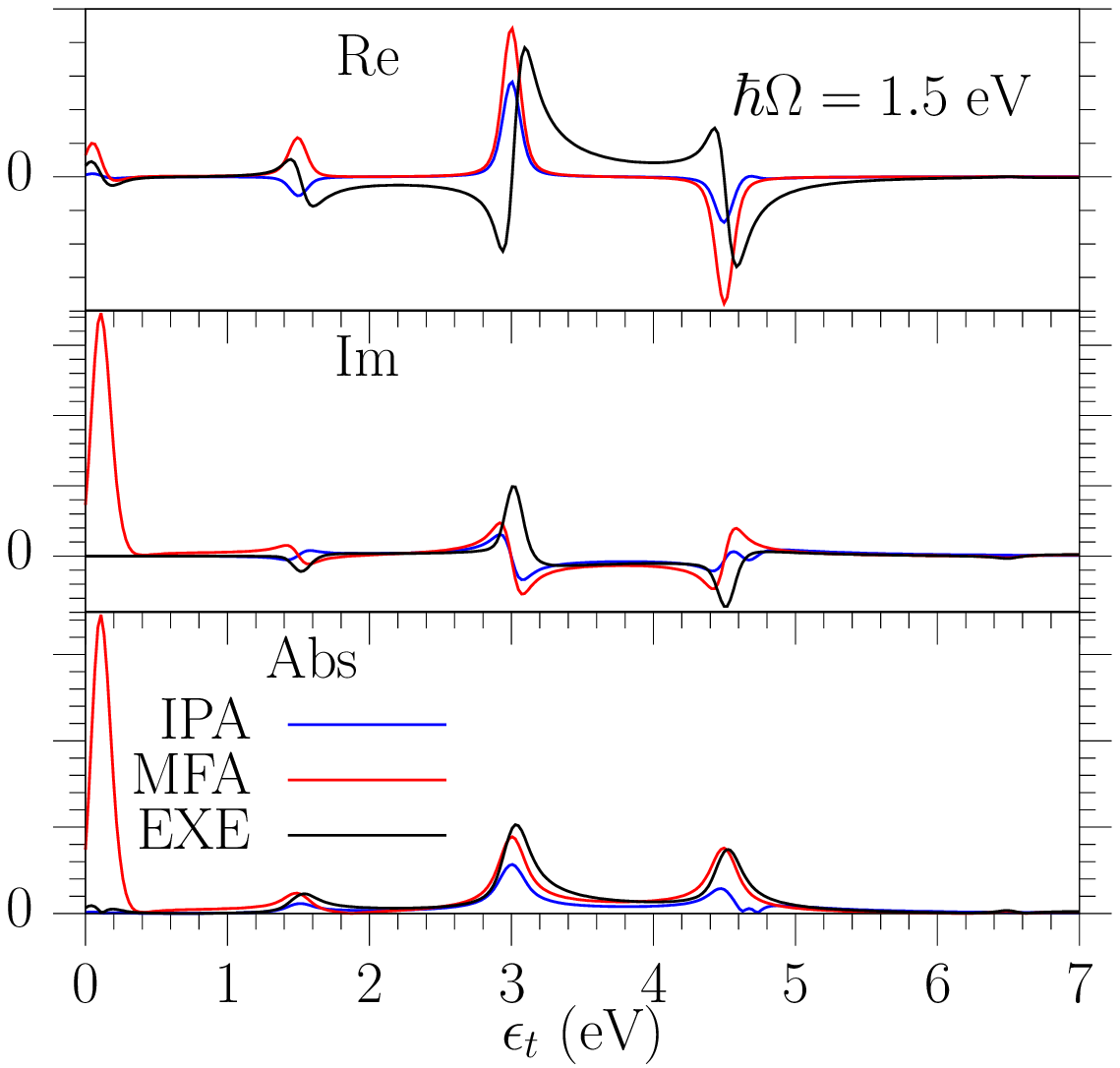}
  \includegraphics[height=5cm]{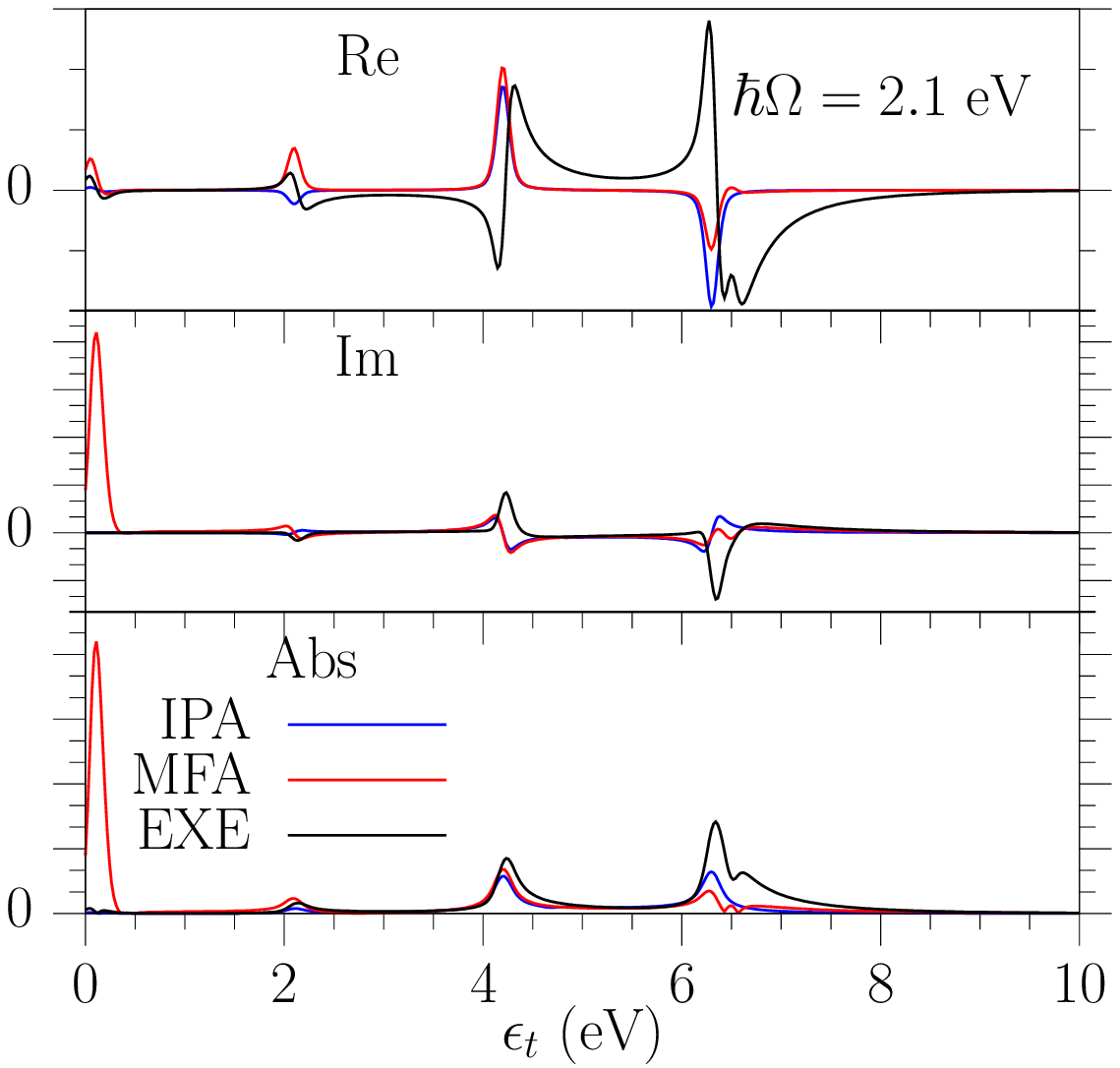}  
\caption{Transition energy resolved THG conductivity for fundamental photon
energies $\hbar\Omega=0.7$~eV, 1.5~eV, and 2.1~eV. }
\label{fig:thgspw} 
\end{figure*}
\end{widetext}

To gain some insight into  these features, we calculate the
transition-energy-resolved 
THG conductivity, shown in Fig.~\ref{fig:thgspw} for $\hbar\Omega=0.7$~eV,
1.5~eV, and 2.1~eV.  In the single-particle approximations, the
values of $|\sigma^{(3)}(\Omega,\epsilon_{t})|$ show three peak
contributions 
located around transition energies $m\hbar\Omega$ with $m=1,2,3$,
corresponding to the one-photon, two-photon, and three-photon resonant
transitions. In IPA, the two-photon resonant transition destructively
interferes with the one-photon and three-photon resonant transitions  \cite{NewJ.Phys._16_53014_2014_Cheng}
(approximately corresponding to prefactors $-17,64,-45$),
which results in a small THG conductivity. In the MFA, there are of
course changes in the density of states and the dipole matrix elements
that affect the THG, but as well the interference is also greatly
changed due to the modification of the Dirac band structure.  Although
there is still some destructive interference between these transitions,
the sign of the first peak in the MFA is changed compared to that
in the IPA, the cancellation is not complete, and this leads to a
larger THG response. As well, there is an additional peak located
at zero energy. This peak has the same sign as the peak at $2\hbar\Omega$,
and it is larger in the MFA than in the IPA.  With the inclusion of
 excitonic effects,  the transition energy distribution of the absolute
values of THG becomes very similar to that of IPA, especially around
the peak at $\epsilon_{t}=0$; simutaneously, an additional phase
is introduced around each peak that changes both the real and imaginary
parts. The results indicate that the interference between the mean
field contributions and the excitonic effects also exists in the optical
nonlinearity.
For $\hbar\Omega=2.1$~eV, the transition energy
resolved spectra shows that the resonant peaks locate around
$\epsilon_t\sim m \hbar\Omega$, which are similar to the dependence in
the single-particle approximation. This confirms that excitonic effects
play  a minor 
role for the peak around $\hbar\Omega\approx 2.1$~eV in
Fig.~\ref{fig:thg}. 

\subsection{Substrate effects}

\begin{figure}[!htpb]
  \centering
  \includegraphics[height=5.cm]{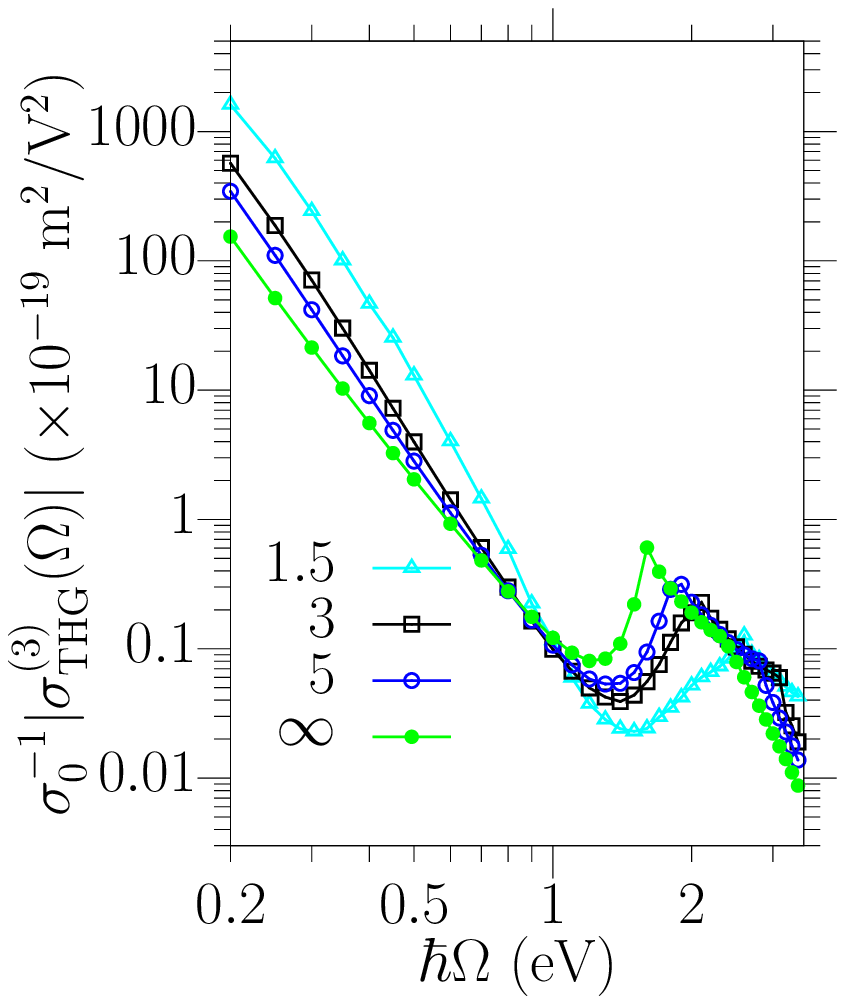}
  \includegraphics[height=5.cm]{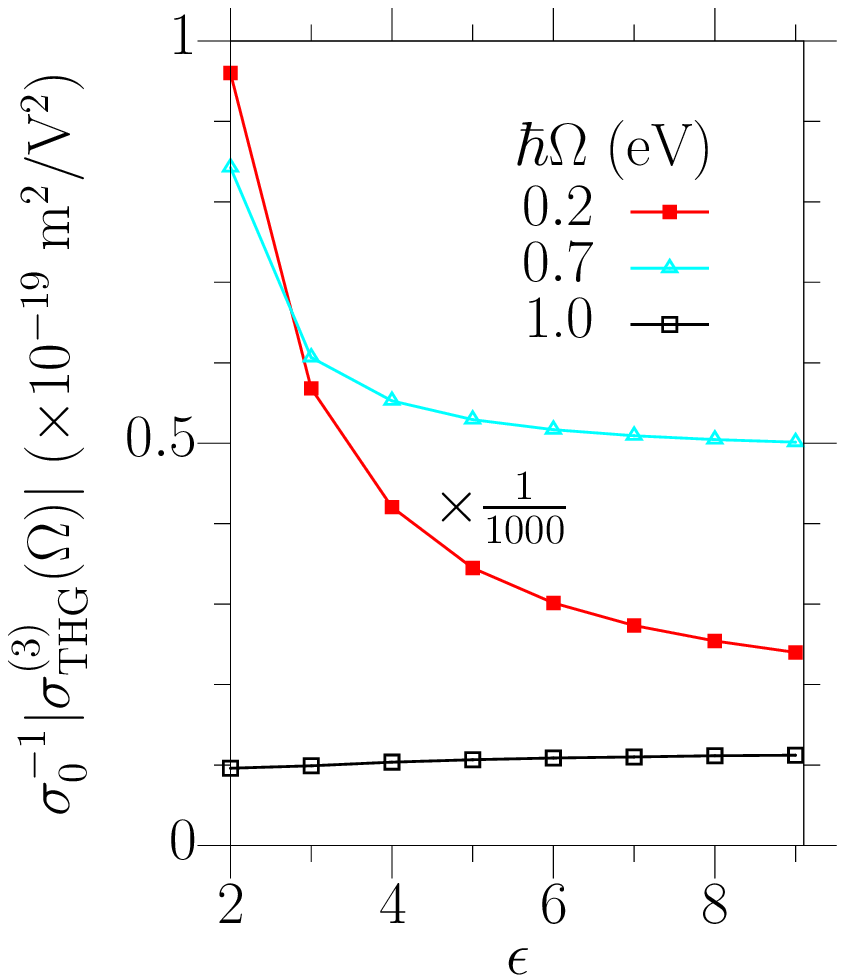}
\caption{Effects of substrate dielectric constant $\epsilon$ on THG conductivity.
(a) The spectrum of THG conductivity for different $\epsilon$, (b)
The $\epsilon$ of THG conductivity for three photon energy $\hbar\Omega=0.2$,
$0.7$, and $1.0$~eV. The results for $0.2$~eV are scaled by $10^{-3}$.}
\label{fig:thghweps} 
\end{figure}

Because of the complexity of  the nonlinear optical response  of
graphene, it is easy  to reveal the consequences of 
  many-body effects by changing the substrate, or more generally the environment, which
is very effective in tuning  the interaction strength of the Coulomb
interaction  \cite{Nanoscale_7_18015_2015_Yadav}.  
In Fig.~\ref{fig:thghweps} we show how the background dielectric
constant $\epsilon$ affects the THG. The Coulomb interaction is inversely
proportional to $\epsilon$, and thus a large background dielectric
constant corresponds to a weak interaction; the limit $\epsilon\to\infty$
corresponds to no Coulomb interaction. 

When $\epsilon$ increases, the THG conductivity decreases for photon
energies $\hbar\Omega$ less than about $0.8$~eV, and the power
index $n$ for $\sigma^{(3)}\propto\Omega^{-n}$ also decreases. But
for photon energies $\hbar\Omega$ larger than about $0.8$~eV, the
THG conductivity increases; the peak corresponding to the resonant
three-photon transition at the $M$ point is shifted to a lower photon
energy, because the energy renormalization decreases with the strength
of the Coulomb interaction. The THG conductivity is more sensitive
to the substrate dielectric constant at lower photon energies, and
it can change over one order of magnitude for $\hbar\Omega=0.2$~eV
when $\epsilon$ changes from $2$ to $\infty$. 

\section{Conclusion\label{sec:con}}

We have theoretically investigated the effects of Coulomb interaction
on third harmonic generation of undoped graphene in the unscreened
Hartree-Fock approximation, with the inclusion of mean field energy
correctionS and excitonic effects. Although there are no bound excitons
formed in gapless graphene, the Coulomb interaction still affects
the third harmonic generation significantly. We find that the  Coulomb
interaction can increase  the amplitude of third harmonic generation
at low photon energy, and decrease it at high photon energy. Despite
the fact that saddle point excitons lead to a large energy shift of
 the resonant peak in linear absorption, they leave  no energy fingerprint
on  the three-photon resonance in third harmonic generation. The underlying
physics can be understood by the inclusion of the Coulomb interaction
step by step. At the level of mean field approximation, the Coulomb
interaction greatly modifies the single particle band structure, which
leads to an enhancement of the third harmonic generation by around
two orders of magnitude compared to the results obtained without
Hartree-Fock term. However, with the full inclusion of the
Hartree-Fock term, this 
large enhancement is largely cancelled, except for the Coulomb-induced
 changes of the power scaling. Therefore, there is a very strong
 interference between the 
contributions from the mean field energy correction and excitonic
effects. We found these effects could be revealed experimentally by 
changing the background dielectric constant, leading to changes both
in the absolute
value of the third harmonic generation coefficient and in its frequency
dependence. The strength of the modifications due to changes in the environment is very sensitive to the
fundamental frequency, and a stronger dependence on the environment
dielectric constant can be found at lower incident photon energy.
Therefore, the strong dependence of third harmonic generation on the
Coulomb interaction  may provide a new optical tool for studying the
many-body effect in graphene.  

For undoped graphene, our results show that the Coulomb interaction
can significantly modify third harmonic generation at low frequencies.
Thus it may be important as well in other nonlinear optical phenomena
involving small frequencies, such as degenerate four wave mixing,
coherent current injection, Kerr effects and two-photon absorption,
current induced second order optical nonlinearity, and jerk current
 \cite{APLPhotonics_4_034201_2019_Cheng}. For doped graphene, where
dynamic screening should be important, the unscreened Hartree-Fock
approximation may be not adequate.   Calculations of the optical
nonlinearity 
in that material exhibit many resonances, arising whenever any  photon
energy involved matches the gap induced by a nonzero chemical potential.
How the Coulomb interaction affects these resonances is  an important
but unexplored problem in understanding the physics of the
optical nonlinearity in graphene.

\acknowledgments This work has been supported by K.C.Wong Education
Foundation Grant No. GJTD-2018-08), Scientific research project of
the Chinese Academy of Sciences Grant No. QYZDB-SSW-SYS038, National
Natural Science Foundation of China Grant No. 11774340 and 61705227.
J.E.S. is supported by the Natural Sciences and Engineering Research
Council of Canada. J.L.C acknowledges valuable discussions with
Prof. K. Shen.

\appendix
\section{The derivation of the equation of motion\label{app:derivation}}

We start with a tight binding Hamiltonian $\hat{H}=\hat{H}_{0}+\hat{H}_{c}+e\bm{E}(t)\cdot\hat{\bm{r}}$
with 
\begin{align}
\hat{H}_{0} & =\sum_{i\alpha,j\beta,\sigma}\gamma_{i-j,\alpha\beta}a_{i\alpha\sigma}^{\dag}a_{j\beta\sigma}\,,\\
\hat{H}_{c} & =\frac{1}{2}\sum_{{i\alpha\sigma_{1}\atop j\beta\sigma_{2}}}V_{i-j,\alpha\beta}a_{i\alpha\sigma_{1}}^{\dag}a_{j\beta\sigma_{2}}^{\dag}a_{j\beta\sigma_{2}}a_{i\alpha\sigma_{1}}\,,\\
\hat{\bm{r}} & =\sum_{i\alpha\sigma}\bm{R}_{i\alpha}a_{i\alpha\sigma}^{\dag}a_{i\alpha\sigma}\,.
\end{align}
Here $a_{i\alpha\sigma}$ is an annihilation operator of an electronic
$p_{z}$ orbital with a spin $\sigma$~($=\uparrow,\downarrow)$
at the site $\bm{R}_{i\alpha}=\bm{R}_{i}+\bm{\tau}_{\alpha}$, where
$i=(n_{i},m_{i})$ is an abbreviated notation for $\bm{R}_{i}=n_{i}\bm{a}_{1}+m_{i}\bm{a}_{2}$.
The term $\hat{H}_{0}$ gives the unperturbated tight binding Hamiltonian.
In this paper, we use the parameters 
\begin{align}
\gamma_{i,AB} & =-\gamma_{0}[\delta_{i,(0,0)}+\delta_{i,(1,0)}+\delta_{i,(0,1)}] \,,\\
 \gamma_{i,BA} & =-\gamma_{0}[\delta_{i,(0,0)}+\delta_{i,(-1,0)}+\delta_{i,(0,-1)}]\,.
\end{align}

We treat the Coulomb interaction $\hat{H}_{c}$  at the level of
Hartree-Fock approximation, and take
\begin{align}
\hat{H}_{c} & \rightarrow\sum_{{i\alpha\sigma_{1}\atop j\beta\sigma_{2}}}V_{i-j,\alpha\beta}\left[\langle a_{i\alpha\sigma_{1}}^{\dag}a_{i\alpha\sigma_{1}}\rangle a_{j\beta\sigma_{2}}^{\dag}a_{j\beta\sigma_{2}}\right.\nonumber \\
 & \left.-\langle a_{i\alpha\sigma_{1}}^{\dag}a_{j\beta\sigma_{2}}\rangle a_{j\beta\sigma_{2}}^{\dag}a_{i\alpha\sigma_{1}}\right]\,.
\end{align}
The constant energy shift in this approximation has been ignored.
The notation $\langle P\rangle$ stands for the statistical average
of the operator $P$ over the ground state. Considering a paramagnetic
ground state in extended graphene, the term $\langle a_{i\alpha\sigma_{1}}^{\dag}a_{j\beta\sigma_{2}}\rangle$
should only be a function of $i-j$ due to the translational symmetry.
Then the first term can be seen to be a simple energy shift. 
By ignoring the spin we get an effective Hamiltonian as 
\begin{align}
\hat{H}_{\text{eff}}(t)=\sum_{i\alpha j\beta}\left[\gamma_{i-j,\alpha\beta}+\overline{H}_{i-j,\alpha\beta}^{\text{HF}}\right]a_{i\alpha}^{\dag}a_{j\beta}+e\bm{E}(t)\cdot\hat{\bm{r}}\,,
\end{align}
with 
\begin{align}
\overline{H}_{i-j,\alpha\beta}^{\text{HF}}=-V_{i-j,\alpha\beta}\langle a_{j\beta}^{\dag}a_{i\alpha}\rangle\,.
\end{align}

In this work, we describe the dynamics of the system by a density
matrix $\overline{\rho}_{i-j,\alpha\beta}(t)=\langle a_{j\beta}^{\dag}(t)a_{i\alpha}(t)\rangle$,
with $a_{i\alpha}(t)$ being the operator $a_{i\alpha}$ in Heisenburg
representation. In the Hartree-Fock approximation, it satisfies the equation
of motion
\begin{align}
i\hbar\partial_{t}\overline{\rho}_{i;\alpha\beta}(t) & =\sum_{j}[\gamma_{i-j}+\overline{H}_{i-j}^{\text{HF}},\overline{\rho}_{j}(t)]_{\alpha\beta}\nonumber \\
 & +e\bm{E}(t)\cdot(\bm{R}_{i}+\bm{\tau}_{\alpha}-\bm{\tau}_{\beta})\overline{\rho}_{i;\alpha\beta}(t)\nonumber \\
 & -i\Gamma[\overline{\rho}_{i;\alpha\beta}(t)-\overline{\rho}_{i;\alpha\beta}^{0}]\,.\label{eq:eom}
\end{align}
In the commutator, $\gamma_{i}$, $H_{i}^{\text{HF}}$, and $\overline{\rho}_{i}(t)$
are treated as matrices with elements determined by the indexes $\alpha\beta$.
The last term models the relaxation process with one phenomenological
 energy parameter $\Gamma$, and $\rho_{i;\alpha\beta}^{0}$
gives the equilibrium distribution including the Hartree-Fock effects.
The velocity operator is given by $\hat{\bm{v}}=(i\hbar)^{-1}[\hat{\bm{r}},\hat{H}_{\text{eff}}(t)]=\sum_{ij\alpha\beta}\bm{v}_{i-j,\alpha\beta}a_{i\alpha}^{\dag}a_{j\beta}$
with 
\begin{align}
\bm{v}_{j,\alpha\beta} & =(i\hbar)^{-1}(\bm{R}_{j}+\bm{\tau}_{\alpha}-\bm{\tau}_{\beta})\left(\gamma_{j,\alpha\beta}+\overline{H}_{j;\alpha\beta}^{\text{HF}}\right)\,.
\end{align}
Then the current density can be calculated through 
\begin{align}
\bm{J}(t)=-e\sum_{j\alpha\beta}\bm{v}_{j,\alpha\beta}\rho_{-j;\beta\alpha}(t)\,.
\end{align}
After Fourier transform of the index $i$ to wave
vectors $\bm{k}$, \textit{i.e.} $\gamma_{i}\to H_{\bm{k}}^{0}$,
$\overline{H}_{j}^{\text{HF}}\to H_{\bm{k}}^{\text{HF}}$, $\overline{\rho}_{i}\to\rho_{\bm{k}}$,
and changing to the moving frame, we get the equations in the main text.


\begin{thebibliography}{42}%
\makeatletter
\providecommand \@ifxundefined [1]{%
 \@ifx{#1\undefined}
}%
\providecommand \@ifnum [1]{%
 \ifnum #1\expandafter \@firstoftwo
 \else \expandafter \@secondoftwo
 \fi
}%
\providecommand \@ifx [1]{%
 \ifx #1\expandafter \@firstoftwo
 \else \expandafter \@secondoftwo
 \fi
}%
\providecommand \natexlab [1]{#1}%
\providecommand \enquote  [1]{``#1''}%
\providecommand \bibnamefont  [1]{#1}%
\providecommand \bibfnamefont [1]{#1}%
\providecommand \citenamefont [1]{#1}%
\providecommand \href@noop [0]{\@secondoftwo}%
\providecommand \href [0]{\begingroup \@sanitize@url \@href}%
\providecommand \@href[1]{\@@startlink{#1}\@@href}%
\providecommand \@@href[1]{\endgroup#1\@@endlink}%
\providecommand \@sanitize@url [0]{\catcode `\\12\catcode `\$12\catcode
  `\&12\catcode `\#12\catcode `\^12\catcode `\_12\catcode `\%12\relax}%
\providecommand \@@startlink[1]{}%
\providecommand \@@endlink[0]{}%
\providecommand \url  [0]{\begingroup\@sanitize@url \@url }%
\providecommand \@url [1]{\endgroup\@href {#1}{\urlprefix }}%
\providecommand \urlprefix  [0]{URL }%
\providecommand \Eprint [0]{\href }%
\providecommand \doibase [0]{http://dx.doi.org/}%
\providecommand \selectlanguage [0]{\@gobble}%
\providecommand \bibinfo  [0]{\@secondoftwo}%
\providecommand \bibfield  [0]{\@secondoftwo}%
\providecommand \translation [1]{[#1]}%
\providecommand \BibitemOpen [0]{}%
\providecommand \bibitemStop [0]{}%
\providecommand \bibitemNoStop [0]{.\EOS\space}%
\providecommand \EOS [0]{\spacefactor3000\relax}%
\providecommand \BibitemShut  [1]{\csname bibitem#1\endcsname}%
\let\auto@bib@innerbib\@empty
\bibitem [{\citenamefont {Kira}\ and\ \citenamefont
  {Koch}(2006)}]{Prog.QuantumElectron._30_155_2006_Kira}%
  \BibitemOpen
  \bibfield  {author} {\bibinfo {author} {\bibfnamefont {M.}~\bibnamefont
  {Kira}}\ and\ \bibinfo {author} {\bibfnamefont {S.}~\bibnamefont {Koch}},\
  }\href {\doibase 10.1016/j.pquantelec.2006.12.002} {\bibfield  {journal}
  {\bibinfo  {journal} {Prog. Quantum Electron.}\ }\textbf {\bibinfo {volume}
  {30}},\ \bibinfo {pages} {155} (\bibinfo {year} {2006})}\BibitemShut
  {NoStop}%
\bibitem [{DFT()}]{DFT_GW_Louie1}%
  \BibitemOpen
  \href {\doibase 10.1016/s1572-0934(06)02002-6} {}\bibinfo {note} {S.G. Louie,
  {\it "Chapter 2 Predicting Materials and Properties: Theory of the Ground and
  Excited State"}, in {\it "Conceptual Foundations of Materials - A Standard
  Model for Ground- and Excited-State Properties"} (Elsevier, 2006) pp.
  9-53.}\BibitemShut {Stop}%
\bibitem [{\citenamefont {Hwang}\ and\ \citenamefont
  {Das~Sarma}(2008)}]{Phys.Rev.B_77_081412_2008_Hwang}%
  \BibitemOpen
  \bibfield  {author} {\bibinfo {author} {\bibfnamefont {E.~H.}\ \bibnamefont
  {Hwang}}\ and\ \bibinfo {author} {\bibfnamefont {S.}~\bibnamefont
  {Das~Sarma}},\ }\href {\doibase 10.1103/PhysRevB.77.081412} {\bibfield
  {journal} {\bibinfo  {journal} {Phys. Rev. B}\ }\textbf {\bibinfo {volume}
  {77}},\ \bibinfo {pages} {081412} (\bibinfo {year} {2008})}\BibitemShut
  {NoStop}%
\bibitem [{\citenamefont {Gr{\"o}nqvist}\ \emph {et~al.}(2012)\citenamefont
  {Gr{\"o}nqvist}, \citenamefont {Stroucken}, \citenamefont {Lindberg},\ and\
  \citenamefont {Koch}}]{Euro.Phys.J.B_85_395_2012_Groenqvist}%
  \BibitemOpen
  \bibfield  {author} {\bibinfo {author} {\bibfnamefont {J.}~\bibnamefont
  {Gr{\"o}nqvist}}, \bibinfo {author} {\bibfnamefont {T.}~\bibnamefont
  {Stroucken}}, \bibinfo {author} {\bibfnamefont {M.}~\bibnamefont {Lindberg}},
  \ and\ \bibinfo {author} {\bibfnamefont {S.}~\bibnamefont {Koch}},\ }\href
  {\doibase 10.1140/epjb/e2012-30593-0} {\bibfield  {journal} {\bibinfo
  {journal} {Euro. Phys. J. B}\ }\textbf {\bibinfo {volume} {85}},\ \bibinfo
  {pages} {395} (\bibinfo {year} {2012})}\BibitemShut {NoStop}%
\bibitem [{\citenamefont {Yadav}\ \emph {et~al.}(2015)\citenamefont {Yadav},
  \citenamefont {Srivastava},\ and\ \citenamefont
  {Ghosh}}]{Nanoscale_7_18015_2015_Yadav}%
  \BibitemOpen
  \bibfield  {author} {\bibinfo {author} {\bibfnamefont {P.}~\bibnamefont
  {Yadav}}, \bibinfo {author} {\bibfnamefont {P.~K.}\ \bibnamefont
  {Srivastava}}, \ and\ \bibinfo {author} {\bibfnamefont {S.}~\bibnamefont
  {Ghosh}},\ }\href {\doibase 10.1039/c5nr04800a} {\bibfield  {journal}
  {\bibinfo  {journal} {Nanoscale}\ }\textbf {\bibinfo {volume} {7}},\ \bibinfo
  {pages} {18015} (\bibinfo {year} {2015})}\BibitemShut {NoStop}%
\bibitem [{\citenamefont {Yang}\ \emph {et~al.}(2009)\citenamefont {Yang},
  \citenamefont {Deslippe}, \citenamefont {Park}, \citenamefont {Cohen},\ and\
  \citenamefont {Louie}}]{Phys.Rev.Lett._103_186802_2009_Yang}%
  \BibitemOpen
  \bibfield  {author} {\bibinfo {author} {\bibfnamefont {L.}~\bibnamefont
  {Yang}}, \bibinfo {author} {\bibfnamefont {J.}~\bibnamefont {Deslippe}},
  \bibinfo {author} {\bibfnamefont {C.-H.}\ \bibnamefont {Park}}, \bibinfo
  {author} {\bibfnamefont {M.~L.}\ \bibnamefont {Cohen}}, \ and\ \bibinfo
  {author} {\bibfnamefont {S.~G.}\ \bibnamefont {Louie}},\ }\href {\doibase
  10.1103/PhysRevLett.103.186802} {\bibfield  {journal} {\bibinfo  {journal}
  {Phys. Rev. Lett.}\ }\textbf {\bibinfo {volume} {103}},\ \bibinfo {pages}
  {186802} (\bibinfo {year} {2009})}\BibitemShut {NoStop}%
\bibitem [{\citenamefont {Jung}\ and\ \citenamefont
  {MacDonald}(2011)}]{Phys.Rev.B_84_85446_2011_Jung}%
  \BibitemOpen
  \bibfield  {author} {\bibinfo {author} {\bibfnamefont {J.}~\bibnamefont
  {Jung}}\ and\ \bibinfo {author} {\bibfnamefont {A.~H.}\ \bibnamefont
  {MacDonald}},\ }\href {\doibase 10.1103/physrevb.84.085446} {\bibfield
  {journal} {\bibinfo  {journal} {Phys. Rev. B}\ }\textbf {\bibinfo {volume}
  {84}},\ \bibinfo {pages} {085446} (\bibinfo {year} {2011})}\BibitemShut
  {NoStop}%
\bibitem [{\citenamefont
  {Katsnelson}(2008)}]{Europhys.Lett._84_37001_2008_Katsnelson}%
  \BibitemOpen
  \bibfield  {author} {\bibinfo {author} {\bibfnamefont {M.~I.}\ \bibnamefont
  {Katsnelson}},\ }\href {\doibase 10.1209/0295-5075/84/37001} {\bibfield
  {journal} {\bibinfo  {journal} {Europhys. Lett.}\ }\textbf {\bibinfo {volume}
  {84}},\ \bibinfo {pages} {37001} (\bibinfo {year} {2008})}\BibitemShut
  {NoStop}%
\bibitem [{\citenamefont {Mak}\ \emph {et~al.}(2011)\citenamefont {Mak},
  \citenamefont {Shan},\ and\ \citenamefont
  {Heinz}}]{Phys.Rev.Lett._106_046401_2011_Mak}%
  \BibitemOpen
  \bibfield  {author} {\bibinfo {author} {\bibfnamefont {K.~F.}\ \bibnamefont
  {Mak}}, \bibinfo {author} {\bibfnamefont {J.}~\bibnamefont {Shan}}, \ and\
  \bibinfo {author} {\bibfnamefont {T.~F.}\ \bibnamefont {Heinz}},\ }\href
  {\doibase 10.1103/physrevlett.106.046401} {\bibfield  {journal} {\bibinfo
  {journal} {Phys. Rev. Lett.}\ }\textbf {\bibinfo {volume} {106}},\ \bibinfo
  {pages} {046401} (\bibinfo {year} {2011})}\BibitemShut {NoStop}%
\bibitem [{\citenamefont {Mak}\ \emph {et~al.}(2014)\citenamefont {Mak},
  \citenamefont {da~Jornada}, \citenamefont {He}, \citenamefont {Deslippe},
  \citenamefont {Petrone}, \citenamefont {Hone}, \citenamefont {Shan},
  \citenamefont {Louie},\ and\ \citenamefont
  {Heinz}}]{Phys.Rev.Lett._112_207401_2014_Mak}%
  \BibitemOpen
  \bibfield  {author} {\bibinfo {author} {\bibfnamefont {K.~F.}\ \bibnamefont
  {Mak}}, \bibinfo {author} {\bibfnamefont {F.~H.}\ \bibnamefont {da~Jornada}},
  \bibinfo {author} {\bibfnamefont {K.}~\bibnamefont {He}}, \bibinfo {author}
  {\bibfnamefont {J.}~\bibnamefont {Deslippe}}, \bibinfo {author}
  {\bibfnamefont {N.}~\bibnamefont {Petrone}}, \bibinfo {author} {\bibfnamefont
  {J.}~\bibnamefont {Hone}}, \bibinfo {author} {\bibfnamefont {J.}~\bibnamefont
  {Shan}}, \bibinfo {author} {\bibfnamefont {S.~G.}\ \bibnamefont {Louie}}, \
  and\ \bibinfo {author} {\bibfnamefont {T.~F.}\ \bibnamefont {Heinz}},\ }\href
  {\doibase 10.1103/physrevlett.112.207401} {\bibfield  {journal} {\bibinfo
  {journal} {Phys. Rev. Lett.}\ }\textbf {\bibinfo {volume} {112}},\ \bibinfo
  {pages} {207401} (\bibinfo {year} {2014})}\BibitemShut {NoStop}%
\bibitem [{\citenamefont {Malic}\ \emph {et~al.}(2011)\citenamefont {Malic},
  \citenamefont {Winzer}, \citenamefont {Bobkin},\ and\ \citenamefont
  {Knorr}}]{Phys.Rev.B_84_205406_2011_Malic}%
  \BibitemOpen
  \bibfield  {author} {\bibinfo {author} {\bibfnamefont {E.}~\bibnamefont
  {Malic}}, \bibinfo {author} {\bibfnamefont {T.}~\bibnamefont {Winzer}},
  \bibinfo {author} {\bibfnamefont {E.}~\bibnamefont {Bobkin}}, \ and\ \bibinfo
  {author} {\bibfnamefont {A.}~\bibnamefont {Knorr}},\ }\href {\doibase
  10.1103/physrevb.84.205406} {\bibfield  {journal} {\bibinfo  {journal} {Phys.
  Rev. B}\ }\textbf {\bibinfo {volume} {84}},\ \bibinfo {pages} {205406}
  (\bibinfo {year} {2011})}\BibitemShut {NoStop}%
\bibitem [{\citenamefont {Avetissian}\ and\ \citenamefont
  {Mkrtchian}(2018)}]{Phys.Rev.B_97_115454_2018_Avetissian}%
  \BibitemOpen
  \bibfield  {author} {\bibinfo {author} {\bibfnamefont {H.~K.}\ \bibnamefont
  {Avetissian}}\ and\ \bibinfo {author} {\bibfnamefont {G.~F.}\ \bibnamefont
  {Mkrtchian}},\ }\href {\doibase 10.1103/PhysRevB.97.115454} {\bibfield
  {journal} {\bibinfo  {journal} {Phys. Rev. B}\ }\textbf {\bibinfo {volume}
  {97}},\ \bibinfo {pages} {115454} (\bibinfo {year} {2018})}\BibitemShut
  {NoStop}%
\bibitem [{\citenamefont {Zhang}\ \emph {et~al.}(2019)\citenamefont {Zhang},
  \citenamefont {Huang}, \citenamefont {Shan}, \citenamefont {Jiang},
  \citenamefont {Zhang}, \citenamefont {Liu}, \citenamefont {Shi},
  \citenamefont {Cheng}, \citenamefont {Sipe}, \citenamefont {Liu},\ and\
  \citenamefont {Wu}}]{Phys.Rev.Lett._122_047401_2019_Zhang}%
  \BibitemOpen
  \bibfield  {author} {\bibinfo {author} {\bibfnamefont {Y.}~\bibnamefont
  {Zhang}}, \bibinfo {author} {\bibfnamefont {D.}~\bibnamefont {Huang}},
  \bibinfo {author} {\bibfnamefont {Y.}~\bibnamefont {Shan}}, \bibinfo {author}
  {\bibfnamefont {T.}~\bibnamefont {Jiang}}, \bibinfo {author} {\bibfnamefont
  {Z.}~\bibnamefont {Zhang}}, \bibinfo {author} {\bibfnamefont
  {K.}~\bibnamefont {Liu}}, \bibinfo {author} {\bibfnamefont {L.}~\bibnamefont
  {Shi}}, \bibinfo {author} {\bibfnamefont {J.}~\bibnamefont {Cheng}}, \bibinfo
  {author} {\bibfnamefont {J.~E.}\ \bibnamefont {Sipe}}, \bibinfo {author}
  {\bibfnamefont {W.-T.}\ \bibnamefont {Liu}}, \ and\ \bibinfo {author}
  {\bibfnamefont {S.}~\bibnamefont {Wu}},\ }\href {\doibase
  10.1103/PhysRevLett.122.047401} {\bibfield  {journal} {\bibinfo  {journal}
  {Phys. Rev. Lett.}\ }\textbf {\bibinfo {volume} {122}},\ \bibinfo {pages}
  {047401} (\bibinfo {year} {2019})}\BibitemShut {NoStop}%
\bibitem [{\citenamefont {Jiang}\ \emph {et~al.}(2018)\citenamefont {Jiang},
  \citenamefont {Huang}, \citenamefont {Cheng}, \citenamefont {Fan},
  \citenamefont {Zhang}, \citenamefont {Shan}, \citenamefont {Yi},
  \citenamefont {Dai}, \citenamefont {Shi}, \citenamefont {Liu}, \citenamefont
  {Zeng}, \citenamefont {Zi}, \citenamefont {Sipe}, \citenamefont {Shen},
  \citenamefont {Liu},\ and\ \citenamefont {Wu}}]{Nat.Photon.___2018_Jiang}%
  \BibitemOpen
  \bibfield  {author} {\bibinfo {author} {\bibfnamefont {T.}~\bibnamefont
  {Jiang}}, \bibinfo {author} {\bibfnamefont {D.}~\bibnamefont {Huang}},
  \bibinfo {author} {\bibfnamefont {J.}~\bibnamefont {Cheng}}, \bibinfo
  {author} {\bibfnamefont {X.}~\bibnamefont {Fan}}, \bibinfo {author}
  {\bibfnamefont {Z.}~\bibnamefont {Zhang}}, \bibinfo {author} {\bibfnamefont
  {Y.}~\bibnamefont {Shan}}, \bibinfo {author} {\bibfnamefont {Y.}~\bibnamefont
  {Yi}}, \bibinfo {author} {\bibfnamefont {Y.}~\bibnamefont {Dai}}, \bibinfo
  {author} {\bibfnamefont {L.}~\bibnamefont {Shi}}, \bibinfo {author}
  {\bibfnamefont {K.}~\bibnamefont {Liu}}, \bibinfo {author} {\bibfnamefont
  {C.}~\bibnamefont {Zeng}}, \bibinfo {author} {\bibfnamefont {J.}~\bibnamefont
  {Zi}}, \bibinfo {author} {\bibfnamefont {J.~E.}\ \bibnamefont {Sipe}},
  \bibinfo {author} {\bibfnamefont {Y.-R.}\ \bibnamefont {Shen}}, \bibinfo
  {author} {\bibfnamefont {W.-T.}\ \bibnamefont {Liu}}, \ and\ \bibinfo
  {author} {\bibfnamefont {S.}~\bibnamefont {Wu}},\ }\href
  {https://doi.org/10.1038/s41566-018-0175-7} {\bibfield  {journal} {\bibinfo
  {journal} {Nat. Photon.}\ }\textbf {\bibinfo {volume} {12}},\ \bibinfo
  {pages} {430} (\bibinfo {year} {2018})},\ \bibinfo {note} {and references
  therein}\BibitemShut {NoStop}%
\bibitem [{\citenamefont {Alexander}\ \emph {et~al.}(2017)\citenamefont
  {Alexander}, \citenamefont {Savostianova}, \citenamefont {Mikhailov},
  \citenamefont {Kuyken},\ and\ \citenamefont
  {Thourhout}}]{ACSPhoton._4_3039_2017_Alexander}%
  \BibitemOpen
  \bibfield  {author} {\bibinfo {author} {\bibfnamefont {K.}~\bibnamefont
  {Alexander}}, \bibinfo {author} {\bibfnamefont {N.~A.}\ \bibnamefont
  {Savostianova}}, \bibinfo {author} {\bibfnamefont {S.~A.}\ \bibnamefont
  {Mikhailov}}, \bibinfo {author} {\bibfnamefont {B.}~\bibnamefont {Kuyken}}, \
  and\ \bibinfo {author} {\bibfnamefont {D.~V.}\ \bibnamefont {Thourhout}},\
  }\href {\doibase 10.1021/acsphotonics.7b00559} {\bibfield  {journal}
  {\bibinfo  {journal} {ACS Photon.}\ }\textbf {\bibinfo {volume} {4}},\
  \bibinfo {pages} {3039} (\bibinfo {year} {2017})}\BibitemShut {NoStop}%
\bibitem [{\citenamefont {Soavi}\ \emph {et~al.}(2018)\citenamefont {Soavi},
  \citenamefont {Wang}, \citenamefont {Rostami}, \citenamefont {Purdie},
  \citenamefont {De~Fazio}, \citenamefont {Ma}, \citenamefont {Luo},
  \citenamefont {Wang}, \citenamefont {Ott}, \citenamefont {Yoon},
  \citenamefont {Bourelle}, \citenamefont {Muench}, \citenamefont {Goykhman},
  \citenamefont {Dal~Conte}, \citenamefont {Celebrano}, \citenamefont
  {Tomadin}, \citenamefont {Polini}, \citenamefont {Cerullo},\ and\
  \citenamefont {Ferrari}}]{Nat.Nano._X_X_2018_Soavi}%
  \BibitemOpen
  \bibfield  {author} {\bibinfo {author} {\bibfnamefont {G.}~\bibnamefont
  {Soavi}}, \bibinfo {author} {\bibfnamefont {G.}~\bibnamefont {Wang}},
  \bibinfo {author} {\bibfnamefont {H.}~\bibnamefont {Rostami}}, \bibinfo
  {author} {\bibfnamefont {D.~G.}\ \bibnamefont {Purdie}}, \bibinfo {author}
  {\bibfnamefont {D.}~\bibnamefont {De~Fazio}}, \bibinfo {author}
  {\bibfnamefont {T.}~\bibnamefont {Ma}}, \bibinfo {author} {\bibfnamefont
  {B.}~\bibnamefont {Luo}}, \bibinfo {author} {\bibfnamefont {J.}~\bibnamefont
  {Wang}}, \bibinfo {author} {\bibfnamefont {A.~K.}\ \bibnamefont {Ott}},
  \bibinfo {author} {\bibfnamefont {D.}~\bibnamefont {Yoon}}, \bibinfo {author}
  {\bibfnamefont {S.~A.}\ \bibnamefont {Bourelle}}, \bibinfo {author}
  {\bibfnamefont {J.~E.}\ \bibnamefont {Muench}}, \bibinfo {author}
  {\bibfnamefont {I.}~\bibnamefont {Goykhman}}, \bibinfo {author}
  {\bibfnamefont {S.}~\bibnamefont {Dal~Conte}}, \bibinfo {author}
  {\bibfnamefont {M.}~\bibnamefont {Celebrano}}, \bibinfo {author}
  {\bibfnamefont {A.}~\bibnamefont {Tomadin}}, \bibinfo {author} {\bibfnamefont
  {M.}~\bibnamefont {Polini}}, \bibinfo {author} {\bibfnamefont
  {G.}~\bibnamefont {Cerullo}}, \ and\ \bibinfo {author} {\bibfnamefont
  {A.~C.}\ \bibnamefont {Ferrari}},\ }\href
  {https://doi.org/10.1038/s41565-018-0145-8} {\bibfield  {journal} {\bibinfo
  {journal} {Nat. Nano.}\ }\textbf {\bibinfo {volume} {13}},\ \bibinfo {pages}
  {583} (\bibinfo {year} {2018})}\BibitemShut {NoStop}%
\bibitem [{\citenamefont {Yoshikawa}\ \emph {et~al.}(2017)\citenamefont
  {Yoshikawa}, \citenamefont {Tamaya},\ and\ \citenamefont
  {Tanaka}}]{Science_356_736_2017_Yoshikawa}%
  \BibitemOpen
  \bibfield  {author} {\bibinfo {author} {\bibfnamefont {N.}~\bibnamefont
  {Yoshikawa}}, \bibinfo {author} {\bibfnamefont {T.}~\bibnamefont {Tamaya}}, \
  and\ \bibinfo {author} {\bibfnamefont {K.}~\bibnamefont {Tanaka}},\ }\href
  {\doibase 10.1126/science.aam8861} {\bibfield  {journal} {\bibinfo  {journal}
  {Science}\ }\textbf {\bibinfo {volume} {356}},\ \bibinfo {pages} {736}
  (\bibinfo {year} {2017})}\BibitemShut {NoStop}%
\bibitem [{\citenamefont {Al-Naib}\ \emph {et~al.}(2014)\citenamefont
  {Al-Naib}, \citenamefont {Sipe},\ and\ \citenamefont
  {Dignam}}]{Phys.Rev.B_90_245423_2014_Al-Naib}%
  \BibitemOpen
  \bibfield  {author} {\bibinfo {author} {\bibfnamefont {I.}~\bibnamefont
  {Al-Naib}}, \bibinfo {author} {\bibfnamefont {J.~E.}\ \bibnamefont {Sipe}}, \
  and\ \bibinfo {author} {\bibfnamefont {M.~M.}\ \bibnamefont {Dignam}},\
  }\href {\doibase 10.1103/physrevb.90.245423} {\bibfield  {journal} {\bibinfo
  {journal} {Phys. Rev. B}\ }\textbf {\bibinfo {volume} {90}},\ \bibinfo
  {pages} {245423} (\bibinfo {year} {2014})}\BibitemShut {NoStop}%
\bibitem [{\citenamefont {Baudisch}\ \emph {et~al.}(2018)\citenamefont
  {Baudisch}, \citenamefont {Marini}, \citenamefont {Cox}, \citenamefont {Zhu},
  \citenamefont {Silva}, \citenamefont {Teichmann}, \citenamefont {Massicotte},
  \citenamefont {Koppens}, \citenamefont {Levitov}, \citenamefont {de~Abajo},\
  and\ \citenamefont {Biegert}}]{Nat.Commun._9_1018_2018_Baudisch}%
  \BibitemOpen
  \bibfield  {author} {\bibinfo {author} {\bibfnamefont {M.}~\bibnamefont
  {Baudisch}}, \bibinfo {author} {\bibfnamefont {A.}~\bibnamefont {Marini}},
  \bibinfo {author} {\bibfnamefont {J.~D.}\ \bibnamefont {Cox}}, \bibinfo
  {author} {\bibfnamefont {T.}~\bibnamefont {Zhu}}, \bibinfo {author}
  {\bibfnamefont {F.}~\bibnamefont {Silva}}, \bibinfo {author} {\bibfnamefont
  {S.}~\bibnamefont {Teichmann}}, \bibinfo {author} {\bibfnamefont
  {M.}~\bibnamefont {Massicotte}}, \bibinfo {author} {\bibfnamefont
  {F.}~\bibnamefont {Koppens}}, \bibinfo {author} {\bibfnamefont {L.~S.}\
  \bibnamefont {Levitov}}, \bibinfo {author} {\bibfnamefont {F.~J.~G.}\
  \bibnamefont {de~Abajo}}, \ and\ \bibinfo {author} {\bibfnamefont
  {J.}~\bibnamefont {Biegert}},\ }\href {\doibase 10.1038/s41467-018-03413-7}
  {\bibfield  {journal} {\bibinfo  {journal} {Nat. Commun.}\ }\textbf {\bibinfo
  {volume} {9}},\ \bibinfo {pages} {1018} (\bibinfo {year} {2018})}\BibitemShut
  {NoStop}%
\bibitem [{\citenamefont {Bonaccorso}\ \emph {et~al.}(2010)\citenamefont
  {Bonaccorso}, \citenamefont {Sun}, \citenamefont {Hasan},\ and\ \citenamefont
  {Ferrari}}]{Nat.Photon._4_611_2010_Bonaccorso}%
  \BibitemOpen
  \bibfield  {author} {\bibinfo {author} {\bibfnamefont {F.}~\bibnamefont
  {Bonaccorso}}, \bibinfo {author} {\bibfnamefont {Z.}~\bibnamefont {Sun}},
  \bibinfo {author} {\bibfnamefont {T.}~\bibnamefont {Hasan}}, \ and\ \bibinfo
  {author} {\bibfnamefont {A.~C.}\ \bibnamefont {Ferrari}},\ }\href@noop {}
  {\bibfield  {journal} {\bibinfo  {journal} {Nat. Photon.}\ }\textbf {\bibinfo
  {volume} {4}},\ \bibinfo {pages} {611} (\bibinfo {year} {2010})}\BibitemShut
  {NoStop}%
\bibitem [{\citenamefont {Sun}\ \emph {et~al.}(2016)\citenamefont {Sun},
  \citenamefont {Martinez},\ and\ \citenamefont
  {Wang}}]{Nat.Photon._10_227_2016_Sun}%
  \BibitemOpen
  \bibfield  {author} {\bibinfo {author} {\bibfnamefont {Z.}~\bibnamefont
  {Sun}}, \bibinfo {author} {\bibfnamefont {A.}~\bibnamefont {Martinez}}, \
  and\ \bibinfo {author} {\bibfnamefont {F.}~\bibnamefont {Wang}},\ }\href
  {\doibase 10.1038/nphoton.2016.15} {\bibfield  {journal} {\bibinfo  {journal}
  {Nat. Photon.}\ }\textbf {\bibinfo {volume} {10}},\ \bibinfo {pages} {227}
  (\bibinfo {year} {2016})}\BibitemShut {NoStop}%
\bibitem [{\citenamefont {Autere}\ \emph {et~al.}(2018)\citenamefont {Autere},
  \citenamefont {Jussila}, \citenamefont {Dai}, \citenamefont {Wang},
  \citenamefont {Lipsanen},\ and\ \citenamefont
  {Sun}}]{Adv.Mater._30_1705963_2018_Autere}%
  \BibitemOpen
  \bibfield  {author} {\bibinfo {author} {\bibfnamefont {A.}~\bibnamefont
  {Autere}}, \bibinfo {author} {\bibfnamefont {H.}~\bibnamefont {Jussila}},
  \bibinfo {author} {\bibfnamefont {Y.}~\bibnamefont {Dai}}, \bibinfo {author}
  {\bibfnamefont {Y.}~\bibnamefont {Wang}}, \bibinfo {author} {\bibfnamefont
  {H.}~\bibnamefont {Lipsanen}}, \ and\ \bibinfo {author} {\bibfnamefont
  {Z.}~\bibnamefont {Sun}},\ }\href {\doibase 10.1002/adma.201705963}
  {\bibfield  {journal} {\bibinfo  {journal} {Adv. Mater.}\ }\textbf {\bibinfo
  {volume} {30}},\ \bibinfo {pages} {1705963} (\bibinfo {year}
  {2018})}\BibitemShut {NoStop}%
\bibitem [{\citenamefont
  {Mikhailov}(2007)}]{Europhys.Lett._79_27002_2007_Mikhailov}%
  \BibitemOpen
  \bibfield  {author} {\bibinfo {author} {\bibfnamefont {S.~A.}\ \bibnamefont
  {Mikhailov}},\ }\href {http://stacks.iop.org/0295-5075/79/i=2/a=27002}
  {\bibfield  {journal} {\bibinfo  {journal} {Europhys. Lett.}\ }\textbf
  {\bibinfo {volume} {79}},\ \bibinfo {pages} {27002} (\bibinfo {year}
  {2007})}\BibitemShut {NoStop}%
\bibitem [{\citenamefont {Hendry}\ \emph {et~al.}(2010)\citenamefont {Hendry},
  \citenamefont {Hale}, \citenamefont {Moger}, \citenamefont {Savchenko},\ and\
  \citenamefont {Mikhailov}}]{Phys.Rev.Lett._105_097401_2010_Hendry}%
  \BibitemOpen
  \bibfield  {author} {\bibinfo {author} {\bibfnamefont {E.}~\bibnamefont
  {Hendry}}, \bibinfo {author} {\bibfnamefont {P.~J.}\ \bibnamefont {Hale}},
  \bibinfo {author} {\bibfnamefont {J.}~\bibnamefont {Moger}}, \bibinfo
  {author} {\bibfnamefont {A.~K.}\ \bibnamefont {Savchenko}}, \ and\ \bibinfo
  {author} {\bibfnamefont {S.~A.}\ \bibnamefont {Mikhailov}},\ }\href {\doibase
  10.1103/PhysRevLett.105.097401} {\bibfield  {journal} {\bibinfo  {journal}
  {Phys. Rev. Lett.}\ }\textbf {\bibinfo {volume} {105}},\ \bibinfo {pages}
  {097401} (\bibinfo {year} {2010})}\BibitemShut {NoStop}%
\bibitem [{\citenamefont {Gu}\ \emph {et~al.}(2012)\citenamefont {Gu},
  \citenamefont {Petrone}, \citenamefont {McMillan}, \citenamefont {van~der
  Zande}, \citenamefont {Yu}, \citenamefont {Lo}, \citenamefont {Kwong},
  \citenamefont {Hone},\ and\ \citenamefont
  {Wong}}]{Nat.Photon._6_554_2012_Gu}%
  \BibitemOpen
  \bibfield  {author} {\bibinfo {author} {\bibfnamefont {T.}~\bibnamefont
  {Gu}}, \bibinfo {author} {\bibfnamefont {N.}~\bibnamefont {Petrone}},
  \bibinfo {author} {\bibfnamefont {J.~F.}\ \bibnamefont {McMillan}}, \bibinfo
  {author} {\bibfnamefont {A.}~\bibnamefont {van~der Zande}}, \bibinfo {author}
  {\bibfnamefont {M.}~\bibnamefont {Yu}}, \bibinfo {author} {\bibfnamefont
  {G.~Q.}\ \bibnamefont {Lo}}, \bibinfo {author} {\bibfnamefont {D.~L.}\
  \bibnamefont {Kwong}}, \bibinfo {author} {\bibfnamefont {J.}~\bibnamefont
  {Hone}}, \ and\ \bibinfo {author} {\bibfnamefont {C.~W.}\ \bibnamefont
  {Wong}},\ }\href {http://dx.doi.org/10.1038/nphoton.2012.147} {\bibfield
  {journal} {\bibinfo  {journal} {Nat. Photon.}\ }\textbf {\bibinfo {volume}
  {6}},\ \bibinfo {pages} {554} (\bibinfo {year} {2012})}\BibitemShut {NoStop}%
\bibitem [{\citenamefont {Vermeulen}\ \emph {et~al.}(2016)\citenamefont
  {Vermeulen}, \citenamefont {Castell\'o-Lurbe}, \citenamefont {Cheng},
  \citenamefont {Pasternak}, \citenamefont {Krajewska}, \citenamefont {Ciuk},
  \citenamefont {Strupinski}, \citenamefont {Thienpont},\ and\ \citenamefont
  {Van~Erps}}]{Phys.Rev.Appl._6_044006_2016_Vermeulen}%
  \BibitemOpen
  \bibfield  {author} {\bibinfo {author} {\bibfnamefont {N.}~\bibnamefont
  {Vermeulen}}, \bibinfo {author} {\bibfnamefont {D.}~\bibnamefont
  {Castell\'o-Lurbe}}, \bibinfo {author} {\bibfnamefont {J.}~\bibnamefont
  {Cheng}}, \bibinfo {author} {\bibfnamefont {I.}~\bibnamefont {Pasternak}},
  \bibinfo {author} {\bibfnamefont {A.}~\bibnamefont {Krajewska}}, \bibinfo
  {author} {\bibfnamefont {T.}~\bibnamefont {Ciuk}}, \bibinfo {author}
  {\bibfnamefont {W.}~\bibnamefont {Strupinski}}, \bibinfo {author}
  {\bibfnamefont {H.}~\bibnamefont {Thienpont}}, \ and\ \bibinfo {author}
  {\bibfnamefont {J.}~\bibnamefont {Van~Erps}},\ }\href {\doibase
  10.1103/physrevapplied.6.044006} {\bibfield  {journal} {\bibinfo  {journal}
  {Phys. Rev. Appl.}\ }\textbf {\bibinfo {volume} {6}},\ \bibinfo {pages}
  {044006} (\bibinfo {year} {2016})}\BibitemShut {NoStop}%
\bibitem [{\citenamefont {Vermeulen}\ \emph {et~al.}(2018)\citenamefont
  {Vermeulen}, \citenamefont {Castell\'o-Lurbe}, \citenamefont {Khoder},
  \citenamefont {Pasternak}, \citenamefont {Krajewska}, \citenamefont {Ciuk},
  \citenamefont {Strupinski}, \citenamefont {Cheng}, \citenamefont
  {Thienpont},\ and\ \citenamefont
  {Van~Erps}}]{Nat.Commun._9_2675_2018_Vermeulen}%
  \BibitemOpen
  \bibfield  {author} {\bibinfo {author} {\bibfnamefont {N.}~\bibnamefont
  {Vermeulen}}, \bibinfo {author} {\bibfnamefont {D.}~\bibnamefont
  {Castell\'o-Lurbe}}, \bibinfo {author} {\bibfnamefont {M.}~\bibnamefont
  {Khoder}}, \bibinfo {author} {\bibfnamefont {I.}~\bibnamefont {Pasternak}},
  \bibinfo {author} {\bibfnamefont {A.}~\bibnamefont {Krajewska}}, \bibinfo
  {author} {\bibfnamefont {T.}~\bibnamefont {Ciuk}}, \bibinfo {author}
  {\bibfnamefont {W.}~\bibnamefont {Strupinski}}, \bibinfo {author}
  {\bibfnamefont {J.}~\bibnamefont {Cheng}}, \bibinfo {author} {\bibfnamefont
  {H.}~\bibnamefont {Thienpont}}, \ and\ \bibinfo {author} {\bibfnamefont
  {J.}~\bibnamefont {Van~Erps}},\ }\href
  {https://doi.org/10.1038/s41467-018-05081-z} {\bibfield  {journal} {\bibinfo
  {journal} {Nat. Commun.}\ }\textbf {\bibinfo {volume} {9}},\ \bibinfo {pages}
  {2675} (\bibinfo {year} {2018})}\BibitemShut {NoStop}%
\bibitem [{\citenamefont {Cheng}\ \emph {et~al.}(2014)\citenamefont {Cheng},
  \citenamefont {Vermeulen},\ and\ \citenamefont
  {Sipe}}]{NewJ.Phys._16_53014_2014_Cheng}%
  \BibitemOpen
  \bibfield  {author} {\bibinfo {author} {\bibfnamefont {J.~L.}\ \bibnamefont
  {Cheng}}, \bibinfo {author} {\bibfnamefont {N.}~\bibnamefont {Vermeulen}}, \
  and\ \bibinfo {author} {\bibfnamefont {J.~E.}\ \bibnamefont {Sipe}},\ }\href
  {\doibase 10.1088/1367-2630/16/5/053014} {\bibfield  {journal} {\bibinfo
  {journal} {New J. Phys.}\ }\textbf {\bibinfo {volume} {16}},\ \bibinfo
  {pages} {053014} (\bibinfo {year} {2014})}\BibitemShut {NoStop}%
\bibitem [{\citenamefont {Cheng}\ \emph
  {et~al.}(2016{\natexlab{a}})\citenamefont {Cheng}, \citenamefont
  {Vermeulen},\ and\ \citenamefont
  {Sipe}}]{Corrigendum_NewJ.Phys._18_29501_2016_Cheng}%
  \BibitemOpen
  \bibfield  {author} {\bibinfo {author} {\bibfnamefont {J.~L.}\ \bibnamefont
  {Cheng}}, \bibinfo {author} {\bibfnamefont {N.}~\bibnamefont {Vermeulen}}, \
  and\ \bibinfo {author} {\bibfnamefont {J.~E.}\ \bibnamefont {Sipe}},\ }\href
  {\doibase 10.1088/1367-2630/18/2/029501} {\bibfield  {journal} {\bibinfo
  {journal} {New J. Phys.}\ }\textbf {\bibinfo {volume} {18}},\ \bibinfo
  {pages} {029501} (\bibinfo {year} {2016}{\natexlab{a}})}\BibitemShut
  {NoStop}%
\bibitem [{\citenamefont {Cheng}\ \emph
  {et~al.}(2015{\natexlab{a}})\citenamefont {Cheng}, \citenamefont
  {Vermeulen},\ and\ \citenamefont {Sipe}}]{Phys.Rev.B_91_235320_2015_Cheng}%
  \BibitemOpen
  \bibfield  {author} {\bibinfo {author} {\bibfnamefont {J.~L.}\ \bibnamefont
  {Cheng}}, \bibinfo {author} {\bibfnamefont {N.}~\bibnamefont {Vermeulen}}, \
  and\ \bibinfo {author} {\bibfnamefont {J.~E.}\ \bibnamefont {Sipe}},\ }\href
  {\doibase http://dx.doi.org/10.1103/PhysRevB.91.235320} {\bibfield  {journal}
  {\bibinfo  {journal} {Phys. Rev. B}\ }\textbf {\bibinfo {volume} {91}},\
  \bibinfo {pages} {235320} (\bibinfo {year} {2015}{\natexlab{a}})}\BibitemShut
  {NoStop}%
\bibitem [{\citenamefont {Cheng}\ \emph
  {et~al.}(2016{\natexlab{b}})\citenamefont {Cheng}, \citenamefont
  {Vermeulen},\ and\ \citenamefont {Sipe}}]{Phys.Rev.B_93_39904_2016_Cheng}%
  \BibitemOpen
  \bibfield  {author} {\bibinfo {author} {\bibfnamefont {J.~L.}\ \bibnamefont
  {Cheng}}, \bibinfo {author} {\bibfnamefont {N.}~\bibnamefont {Vermeulen}}, \
  and\ \bibinfo {author} {\bibfnamefont {J.~E.}\ \bibnamefont {Sipe}},\ }\href
  {\doibase 10.1103/PhysRevB.93.039904} {\bibfield  {journal} {\bibinfo
  {journal} {Phys. Rev. B}\ }\textbf {\bibinfo {volume} {93}},\ \bibinfo
  {pages} {039904} (\bibinfo {year} {2016}{\natexlab{b}})}\BibitemShut
  {NoStop}%
\bibitem [{\citenamefont {Cheng}\ \emph
  {et~al.}(2015{\natexlab{b}})\citenamefont {Cheng}, \citenamefont
  {Vermeulen},\ and\ \citenamefont {Sipe}}]{Phys.Rev.B_92_235307_2015_Cheng}%
  \BibitemOpen
  \bibfield  {author} {\bibinfo {author} {\bibfnamefont {J.~L.}\ \bibnamefont
  {Cheng}}, \bibinfo {author} {\bibfnamefont {N.}~\bibnamefont {Vermeulen}}, \
  and\ \bibinfo {author} {\bibfnamefont {J.~E.}\ \bibnamefont {Sipe}},\ }\href
  {\doibase 10.1103/PhysRevB.92.235307} {\bibfield  {journal} {\bibinfo
  {journal} {Phys. Rev. B}\ }\textbf {\bibinfo {volume} {92}},\ \bibinfo
  {pages} {235307} (\bibinfo {year} {2015}{\natexlab{b}})}\BibitemShut
  {NoStop}%
\bibitem [{\citenamefont
  {Mikhailov}(2016)}]{Phys.Rev.B_93_085403_2016_Mikhailov}%
  \BibitemOpen
  \bibfield  {author} {\bibinfo {author} {\bibfnamefont {S.~A.}\ \bibnamefont
  {Mikhailov}},\ }\href {\doibase 10.1103/PhysRevB.93.085403} {\bibfield
  {journal} {\bibinfo  {journal} {Phys. Rev. B}\ }\textbf {\bibinfo {volume}
  {93}},\ \bibinfo {pages} {085403} (\bibinfo {year} {2016})}\BibitemShut
  {NoStop}%
\bibitem [{\citenamefont {Jiang}\ \emph {et~al.}(2007)\citenamefont {Jiang},
  \citenamefont {Saito}, \citenamefont {Samsonidze}, \citenamefont {Jorio},
  \citenamefont {Chou}, \citenamefont {Dresselhaus},\ and\ \citenamefont
  {Dresselhaus}}]{Phys.Rev.B_75_035407_2007_Jiang}%
  \BibitemOpen
  \bibfield  {author} {\bibinfo {author} {\bibfnamefont {J.}~\bibnamefont
  {Jiang}}, \bibinfo {author} {\bibfnamefont {R.}~\bibnamefont {Saito}},
  \bibinfo {author} {\bibfnamefont {G.~G.}\ \bibnamefont {Samsonidze}},
  \bibinfo {author} {\bibfnamefont {A.}~\bibnamefont {Jorio}}, \bibinfo
  {author} {\bibfnamefont {S.~G.}\ \bibnamefont {Chou}}, \bibinfo {author}
  {\bibfnamefont {G.}~\bibnamefont {Dresselhaus}}, \ and\ \bibinfo {author}
  {\bibfnamefont {M.~S.}\ \bibnamefont {Dresselhaus}},\ }\href {\doibase
  10.1103/PhysRevB.75.035407} {\bibfield  {journal} {\bibinfo  {journal} {Phys.
  Rev. B}\ }\textbf {\bibinfo {volume} {75}},\ \bibinfo {pages} {035407}
  (\bibinfo {year} {2007})}\BibitemShut {NoStop}%
\bibitem [{Note1()}]{Note1}%
  \BibitemOpen
  \bibinfo {note} {For graphene embedded inside two different materials, the
  background dielectric constant is the average of the dielectric constants.
  Because the interaction is mainly through the electric force outside of the
  graphene plane, the screening from the graphene electrons is
  ignored.}\BibitemShut {Stop}%
\bibitem [{\citenamefont {Hwang}\ \emph {et~al.}(2007)\citenamefont {Hwang},
  \citenamefont {Hu},\ and\ \citenamefont
  {Das~Sarma}}]{Phys.Rev.Lett._99_226801_2007_Hwang}%
  \BibitemOpen
  \bibfield  {author} {\bibinfo {author} {\bibfnamefont {E.~H.}\ \bibnamefont
  {Hwang}}, \bibinfo {author} {\bibfnamefont {B.~Y.-K.}\ \bibnamefont {Hu}}, \
  and\ \bibinfo {author} {\bibfnamefont {S.}~\bibnamefont {Das~Sarma}},\ }\href
  {\doibase 10.1103/physrevlett.99.226801} {\bibfield  {journal} {\bibinfo
  {journal} {Phys. Rev. Lett.}\ }\textbf {\bibinfo {volume} {99}},\ \bibinfo
  {pages} {226801} (\bibinfo {year} {2007})}\BibitemShut {NoStop}%
\bibitem [{\citenamefont {Stroucken}\ \emph {et~al.}(2012)\citenamefont
  {Stroucken}, \citenamefont {Gr\"onqvist},\ and\ \citenamefont
  {Koch}}]{J.Opt.Soc.Am.B_29_A86_2012_Stroucken}%
  \BibitemOpen
  \bibfield  {author} {\bibinfo {author} {\bibfnamefont {T.}~\bibnamefont
  {Stroucken}}, \bibinfo {author} {\bibfnamefont {J.~H.}\ \bibnamefont
  {Gr\"onqvist}}, \ and\ \bibinfo {author} {\bibfnamefont {S.~W.}\ \bibnamefont
  {Koch}},\ }\href {\doibase 10.1364/josab.29.000a86} {\bibfield  {journal}
  {\bibinfo  {journal} {J. Opt. Soc. Am. B}\ }\textbf {\bibinfo {volume}
  {29}},\ \bibinfo {pages} {A86} (\bibinfo {year} {2012})}\BibitemShut
  {NoStop}%
\bibitem [{\citenamefont {Stroucken}\ \emph {et~al.}(2011)\citenamefont
  {Stroucken}, \citenamefont {Gr\"onqvist},\ and\ \citenamefont
  {Koch}}]{Phys.Rev.B_84_205445_2011_Stroucken}%
  \BibitemOpen
  \bibfield  {author} {\bibinfo {author} {\bibfnamefont {T.}~\bibnamefont
  {Stroucken}}, \bibinfo {author} {\bibfnamefont {J.~H.}\ \bibnamefont
  {Gr\"onqvist}}, \ and\ \bibinfo {author} {\bibfnamefont {S.~W.}\ \bibnamefont
  {Koch}},\ }\href {\doibase 10.1103/PhysRevB.84.205445} {\bibfield  {journal}
  {\bibinfo  {journal} {Phys. Rev. B}\ }\textbf {\bibinfo {volume} {84}},\
  \bibinfo {pages} {205445} (\bibinfo {year} {2011})}\BibitemShut {NoStop}%
\bibitem [{Qua()}]{QuantumKineticsinTransportandOpticsofSemiconductors1}%
  \BibitemOpen
  \href@noop {} {}\bibinfo {note} {Hartmut Haug and Antti-Pekka Jauho, {\it
  Quantum Kinetics in Transport and Optics of Semiconductors} (Springer,
  Berlin, 2007).}\BibitemShut {Stop}%
\bibitem [{Note2()}]{Note2}%
  \BibitemOpen
  \bibinfo {note} {This work considers the third harmonic generation in the
  weak field regime, thus the use of a periodic field follows the standard
  perturbative treatment, which should be suitable in experiments where the
  pulse duration is much longer than the relaxation time\cite
  {Phys.Rev.B_92_235307_2015_Cheng}.{}}\BibitemShut {NoStop}%
\bibitem [{\citenamefont {Liu}\ \emph {et~al.}(2017)\citenamefont {Liu},
  \citenamefont {Zhang},\ and\ \citenamefont
  {Cao}}]{Phys.Rev.B_96_035206_2017_Liu}%
  \BibitemOpen
  \bibfield  {author} {\bibinfo {author} {\bibfnamefont {Z.}~\bibnamefont
  {Liu}}, \bibinfo {author} {\bibfnamefont {C.}~\bibnamefont {Zhang}}, \ and\
  \bibinfo {author} {\bibfnamefont {J.~C.}\ \bibnamefont {Cao}},\ }\href
  {\doibase 10.1103/physrevb.96.035206} {\bibfield  {journal} {\bibinfo
  {journal} {Phys. Rev. B}\ }\textbf {\bibinfo {volume} {96}},\ \bibinfo
  {pages} {035206} (\bibinfo {year} {2017})}\BibitemShut {NoStop}%
\bibitem [{\citenamefont {Cheng}\ \emph {et~al.}(2019)\citenamefont {Cheng},
  \citenamefont {Sipe}, \citenamefont {Wu},\ and\ \citenamefont
  {Guo}}]{APLPhotonics_4_034201_2019_Cheng}%
  \BibitemOpen
  \bibfield  {author} {\bibinfo {author} {\bibfnamefont {J.~L.}\ \bibnamefont
  {Cheng}}, \bibinfo {author} {\bibfnamefont {J.~E.}\ \bibnamefont {Sipe}},
  \bibinfo {author} {\bibfnamefont {S.~W.}\ \bibnamefont {Wu}}, \ and\ \bibinfo
  {author} {\bibfnamefont {C.}~\bibnamefont {Guo}},\ }\href {\doibase
  10.1063/1.5053715} {\bibfield  {journal} {\bibinfo  {journal} {{APL}
  Photonics}\ }\textbf {\bibinfo {volume} {4}},\ \bibinfo {pages} {034201}
  (\bibinfo {year} {2019})}\BibitemShut {NoStop}%
\end{thebibliography}

%

\end{document}